\documentclass[usenatbib,usegraphicx]{mn2e}

\usepackage{amsmath}
\usepackage{multirow}

\def\apj{ApJ}
\def\apjs{ApJS}
\def\apjl{ApJL}

\def\mnras{MNRAS}
\def\aj{AJ}
\def\nat{Nature}
\def\araa{ARA\&A}
\def\pasp{PASP}

\def\icarus{Icarus}
\def\ssr{Space Sci. Rev.}

\def\cm{\textrm{cm}}
\def\meter{\textrm{m}}
\def\sec{\textrm{s}}
\def\km{\textrm{km}}
\def\kms{\textrm{km\ s}^{-1}}

\def\pc{\textrm{pc}}

\def\degtext{\textrm{deg}}
\def\muas{\mu\textrm{as}}

\def\msec{\textrm{ms}}
\def\sr{\textrm{sr}}
\def\Ang{\textrm{\AA}}
\def\AU{\textrm{AU}}
\def\Hz{\textrm{Hz}}
\def\Blue{}

\title[Occultations with Cherenkov telescopes]{On the Use of Cherenkov Telescopes for Outer Solar System Body Occultations}
\author[Lacki]{Brian C. Lacki$^{1}$\\$^1$Institute for Advanced Study, Einstein Drive, Princeton, NJ 08540, USA, brianlacki@ias.edu}

\begin{document}
\maketitle

\begin{abstract}
Imaging Atmosphere Cherenkov Telescopes (IACT) are arrays of very large optical telescopes that are well-suited for rapid photometry of bright sources.  I investigate their potential in observing stellar occultations by small objects in the outer Solar System, Transjovian Objects (TJOs).  These occultations cast diffraction patterns on the Earth.  Current IACT arrays are capable of detecting objects smaller than 100 {\Blue metres} in radius in the Kuiper Belt and 1 km radius out to 5000 AU.  The future Cherenkov Telescope Array (CTA) will have even greater capabilities.  Because the arrays include several telescopes, they can potentially measure the speeds of TJOs without degeneracies, and the sizes of the TJOs and background stars.  I estimate the achievable precision using a Fisher matrix analysis.  With CTA, the precisions of these parameter estimations will be as good as a few percent.  I consider how often {\Blue detectable} occultations {\Blue occur} by members of different TJO populations, including Centaurs, Kuiper Belt Objects (KBOs), Oort cloud objects, and satellites and Trojans of Uranus and Neptune.  The great sensitivity of IACT arrays means that they likely detect KBO occultations once every ${\cal O}(10)$ hours when looking near the ecliptic.  IACTs can also set useful limits on many other TJO populations.
\end{abstract}


\section{Introduction}
\label{sec:Introduction}

There are many minor bodies in the Solar System beyond the orbit of Jupiter, the Transjovian Objects (TJOs).  There are several populations of TJOs.  These include the Centaurs, a collection of objects that orbits between Jupiter and Neptune; the Kuiper Belt, a reservoir of bodies such as Pluto orbiting 30 - 50 AU from the Sun; the Scattered Disc Objects, a lower density population (among them, Eris; {\Blue \citealt*{Brown05}; \citealt{Brown07}}) that extends out to 100 - 200 AU {\Blue \citep*[e.g.,][]{Gladman08}}; and the Oort Cloud, a population of comets that is generally believed to mostly reside tens of thousands of AUs from the inner Solar System, but includes objects like Sedna that orbit hundreds of AU away {\Blue \citep*{Brown04}}.  The TJOs formed out of the debris left over after planet formation, and their physical properties potentially contain information about the early Solar System.  Their orbits also record the effects of gravitational perturbations from the planets, and constrain models of Solar System dynamics \citep*[e.g.,][]{Morbidelli08}.  

Because the TJOs are in the distant reaches of the Solar System, our ability to observe and understand them is limited.  Large TJOs can be directly observed with telescopes; following the early discovery of Pluto, other large TJOs were found with surveys starting in the 1990s (e.g., \citealt{Jewitt93}; {\Blue \citealt{Brown04,Brown05,Kavelaars08,Trujillo08}}).  These big objects have been studied intensely with photometry and spectroscopy {\Blue \citep[e.g.,][]{Lazzarin03,Peixinho04,Stransberry08}}.  Pluto itself, {\Blue and possibly} more distant Kuiper Belt objects, will be visited by the \emph{New Horizons} probe, which will increase our understanding of TJOs {\Blue \citep{Stern08}}.\footnote{Triton, the largest moon of Neptune, is likely a captured Kuiper Belt object \citep[e.g.,][]{Agnor06}, and was visited by \emph{Voyager 2} in 1989.  However, its structure was radically altered by its capture \citep*[e.g.,][]{McKinnon95}.}  However, TJOs with radii of a kilometre or less are far more difficult to observe.  General constraints on the amount of mass in small bodies can be found from the level of CMB anisotropies (\citealt*{Babich07}; \citealt{Ichikawa11}), the brightness of the infrared background \citep{Kenyon01}, and the gamma-ray background \citep{Moskalenko08,Moskalenko09}.  

The main method of searching for these {\Blue small} objects is through stellar occultations, when they pass in front of a background star, blocking and diffracting its light (\citealt{Bailey76,Dyson92}; \citealt*{Roques87}; \citealt{Roques00}).  {\Blue At least two TJOs were detected with this method in the past few years \citep{Schlichting09,Schlichting12}.}  The Fresnel scale for objects {\Blue in} the outer Solar System is $\ell_F = \sqrt{\lambda D / 2} = 1.4\ \km\ (\lambda / 5500\ \Ang)^{1/2} (D / 50\ \AU)^{1/2}$.  Objects larger than $\ell_F$ essentially cast a geometrical shadow on the Earth, while smaller objects cast a diffraction pattern on the Earth as wide as $\ell_F$, with the star's magnitude fluctuations set by the size of the occulter.  

Observing these events poses several challenges.  First, the depth of the fluctuations for the smaller TJOs is typically only of order a few percent.  Second, because the Earth moves with a relative speed of up to $\sim 30\ \kms$ with respect to TJOs, the events last only a fraction of a second.  High quality light curves must be therefore sampled with a frequency 5 -- 100 Hz {\Blue (cf., \citealt{Nihei07}, hereafter N07)}.  

Past and present surveys for TJO occultations include TAOS, the Taiwanese American Occultation Survey, which is dedicated to detecting TJO occultations \citep{Lehner09}; additionally, surveys have been carried out on the MMT observatory \citep{Bianco09}, the Panoramic Survey Telescope and Rapid Response System 1 (PS1) telescope \citep{Wang10}, the Very Large Telescope \citep{Doressoundiram13}, and Hubble Space Telescope \citep{Schlichting09}.  Additional searches for occultations of Sco X-1 in the X-ray band have been {\Blue conducted}, but instrumental effects proved troublesome (\citealt{Chang06,Chang07,Jones08,Liu08}; \citealt*{Chang11}).  Historically, a big problem with the occultation method in visible light is scintillation noise, which dominates fluctuations at short frequencies.  The technique therefore requires large telescopes with high frequency sampling, and must deal with scintillation by either having a large signal-to-noise ratio \citep{Wang10,Doressoundiram13}, using multiple telescopes as a veto \citep{Bianco10,Lehner10}, or {\Blue using} space telescopes to eliminate it altogether \citep{Schlichting09}.

Many of these conditions are fulfilled by the Imaging Atmospheric Cherenkov Telescopes (IACTs), which are among the largest optical telescopes in the world.  Cherenkov telescopes are used to detect TeV gamma-rays by imaging the flash of Cherenkov light emitted by particle showers created when the gamma ray hits the upper atmosphere (e.g., {\Blue \citealt{Galbraith53};} \citealt{Aharonian97}).  The flashes are faint, so the Cherenkov telescopes must be large to collect as many photons as possible (see Table~\ref{table:IACTArrays}). In addition, the Cherenkov flashes are very short, only a few nanoseconds long, so Cherenkov telescopes use photomultipliers or silicon photon detectors to sample fluxes on MHz time-scales.  The optical field of view of a Cherenkov telescopes is a few degrees. Finally, Cherenkov telescopes often come in arrays, {\Blue allowing} them to image a particle shower in three dimensions, but {\Blue this} could also be useful for vetoing occultation false positives.  

Cherenkov telescopes achieve this remarkable performance at a low cost by sacrificing angular resolution: the typical {\Blue point spread function (PSF)} of a Cherenkov telescope is a few arcminutes \citep{Bernlohr03,Cornils03}.  This increases the noise of a source because of blending with more sky background.  However, for the brightest sources (with $V \la 8 - 10$), the Poisson noise of the star is greater than the sky background noise, and confusion only becomes a problem at $V \approx 14 - 16$ \citep{Bahcall80}.  Therefore, the Cherenkov telescopes may be very useful {\Blue for observing} bright stars.  Cherenkov telescopes have been used to study the Crab pulsar optical light curve \citep{Hinton06,Lucarelli08}, optical transients on millisecond and microsecond time-scales \citep{Deil09}, and for optical SETI \citep{Eichler01,Holder05}, and they are potentially good for stellar intensity interferometry \citep{LeBohec06}, detecting picosecond optical transients \citep{Borra10} and optical polarimetry \citep{Lacki11}.  

Current Cherenkov telescope arrays include the Very Energetic Radiation Imaging Telescope Array System (VERITAS)\footnote{http://veritas.sao.arizona.edu/} and the High Energy Stereoscopic System (HESS)\footnote{http://www.mpi-hd.mpg.de/hfm/HESS/} (see Table~\ref{table:IACTArrays}).  VERITAS consists of four 12-meter telescopes and is located in the Northern hemisphere \citep{Weekes02}.  HESS is an array of four 12-meter telescopes plus a central, fifth 28-meter telescope, located in the Southern hemisphere \citep{Bernlohr03,Cornils03,Cornils05}.  The power of Cherenkov telescopes will increase much further with the next generation Cherenkov Telescope Array (CTA).\footnote{http://www.cta-observatory.org/}  The array will include four large ($\sim 25\ \meter$) telescopes, as well as tens of medium ($\sim 12\ \meter$) and small ($\sim 7\ \meter$) telescopes \citep{Bernlohr13}.  The sheer number of telescopes will allow a vast number of photons to be collected, and would efficiently eliminate false positives.  

The precise observations of stellar occultations that are possible with Cherenkov telescopes can be useful for studying not only the TJOs themselves but the occulted stars.  The shape of the light curve depends on the size of the TJO, as well as the angular size of the star, in units of the Fresnel scale.  An array of telescopes, by sampling the light curve at different positions, {\Blue could} measure the size of the diffraction pattern, giving the Fresnel scale, from which these parameters and the TJO distance can be calculated.  Besides these parameters, the arrays can measure the position of the TJO on the sky, and the non-radial components of the TJO velocity.  

\begin{table*}
\begin{minipage}{110mm}
\caption{TJO occultation instrument configurations}
\label{table:IACTArrays}
\begin{tabular}{lccccc}
\hline
Array   & $N_{\rm tel}$ & $D_{\rm tel}$ Range & $A_{\rm tel}$ & Baseline Range & References\\
        &               & $m$                 & $m^2$         & $m$            & \\
\hline
TAOS         & 4        & 0.5                 & 0.79          & 7.6 -- 100     & (1)\\
Megacam/MMT  & 1        & 6.5                 & 33            & ...            & (2)\\
ULTRACAM/VLT & 1        & 8.2                 & 53            & ...            & (3)\\
\hline
VERITAS      & 4        & 12.0                & 450           & 83 -- 180      & (4)\\
HESS         & 5        & 12.0 -- 28.0        & 1100          & 85 -- 170      & (5)\\
CTA (E)      & 59       & 7.2 -- 24.0         & 5800          & 67 -- 3500     & (6)\\
\hline
\end{tabular}
\\{\bf References} -- (1) \citet{Lehner09}; (2) -- as used in \citet{Bianco09}; (3) -- as used in \citet{Doressoundiram13}; (4) -- \citet{Weekes02}, present configuration given in \citet*{Perkins09}; (5) -- see \citet{Bernlohr03}, \citet{Cornils05}, and the HESS website, http://www.mpi-hd.mpg.de/hfm/HESS; (6) -- configuration E given in \citet{Bernlohr13}
\\
\end{minipage}
\end{table*}

In \citet{Lacki11}, I computed the expected signal-to-noise ratios for photometric, spectroscopic, and polarimetric measurements by Cherenkov telescopes given the large sky noise.  Noting that these ratios were better for bright sources, I listed several possible phenomena that might be studied optically with Cherenkov telescopes, including stellar occultations by TJOs.  Now I evaluate the abilities of Cherenkov telescopes to detect and characterize these occultations.  After describing how I calculate the light curves of occultation events in Section~\ref{sec:Calculations}, I evaluate the types of events that are detectable at VERITAS, HESS, and CTA in Section~\ref{sec:Detectability}.  The ability of Cherenkov telescopes to estimate the parameters of an occultation is evaluated with Fisher matrix analysis {\Blue and likelihood ratio analysis} in Section~\ref{sec:ParameterEstimation}.  Finally, the frequency that these arrays should observe TJO occultations is calculated in Section~\ref{sec:Frequency}.

\section{Calculation of Light Curves and Derivatives}
\label{sec:Calculations}
\subsection{Observing Geometry}

\begin{figure*}
\centerline{\includegraphics[width=8.5cm]{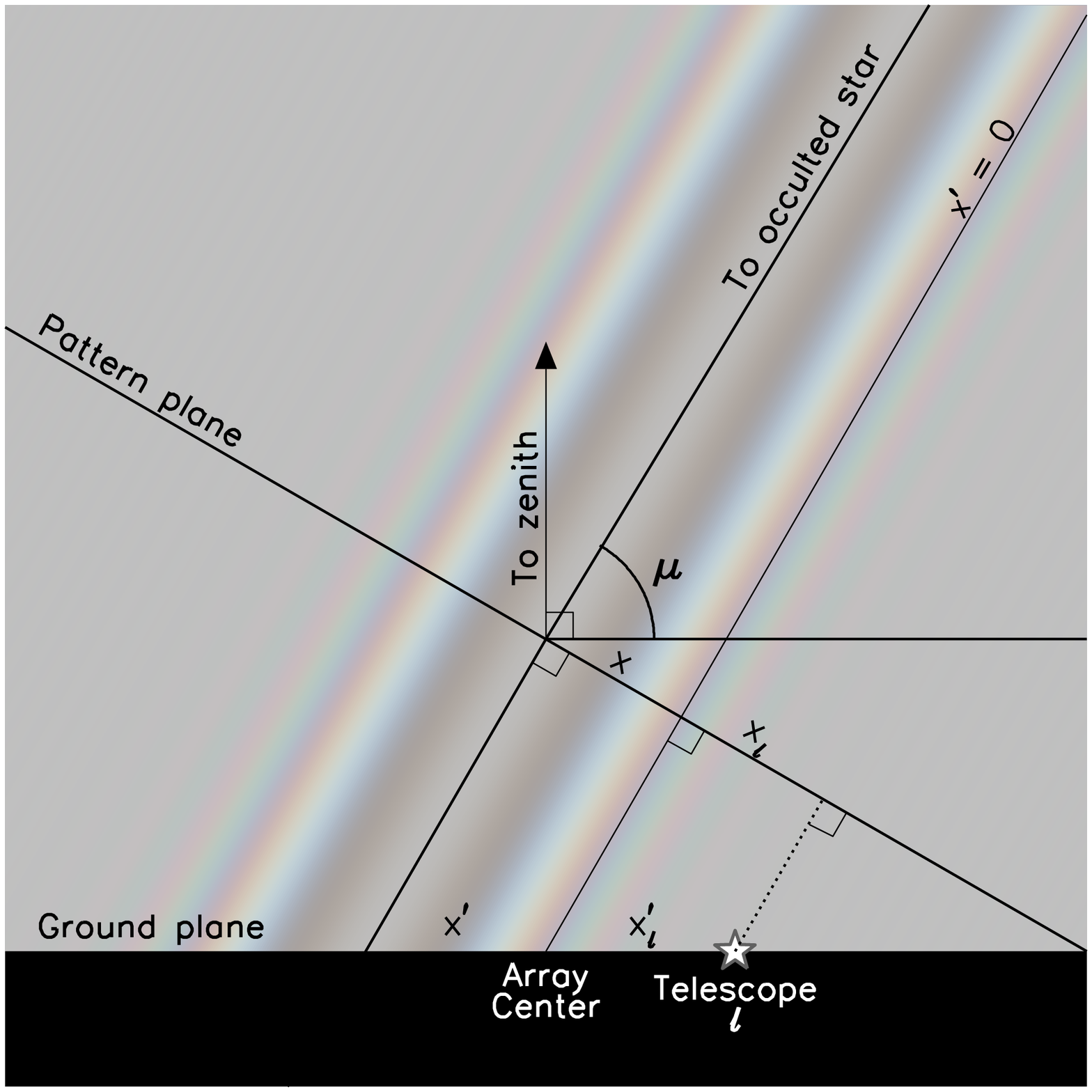}\ \includegraphics[width=8.5cm]{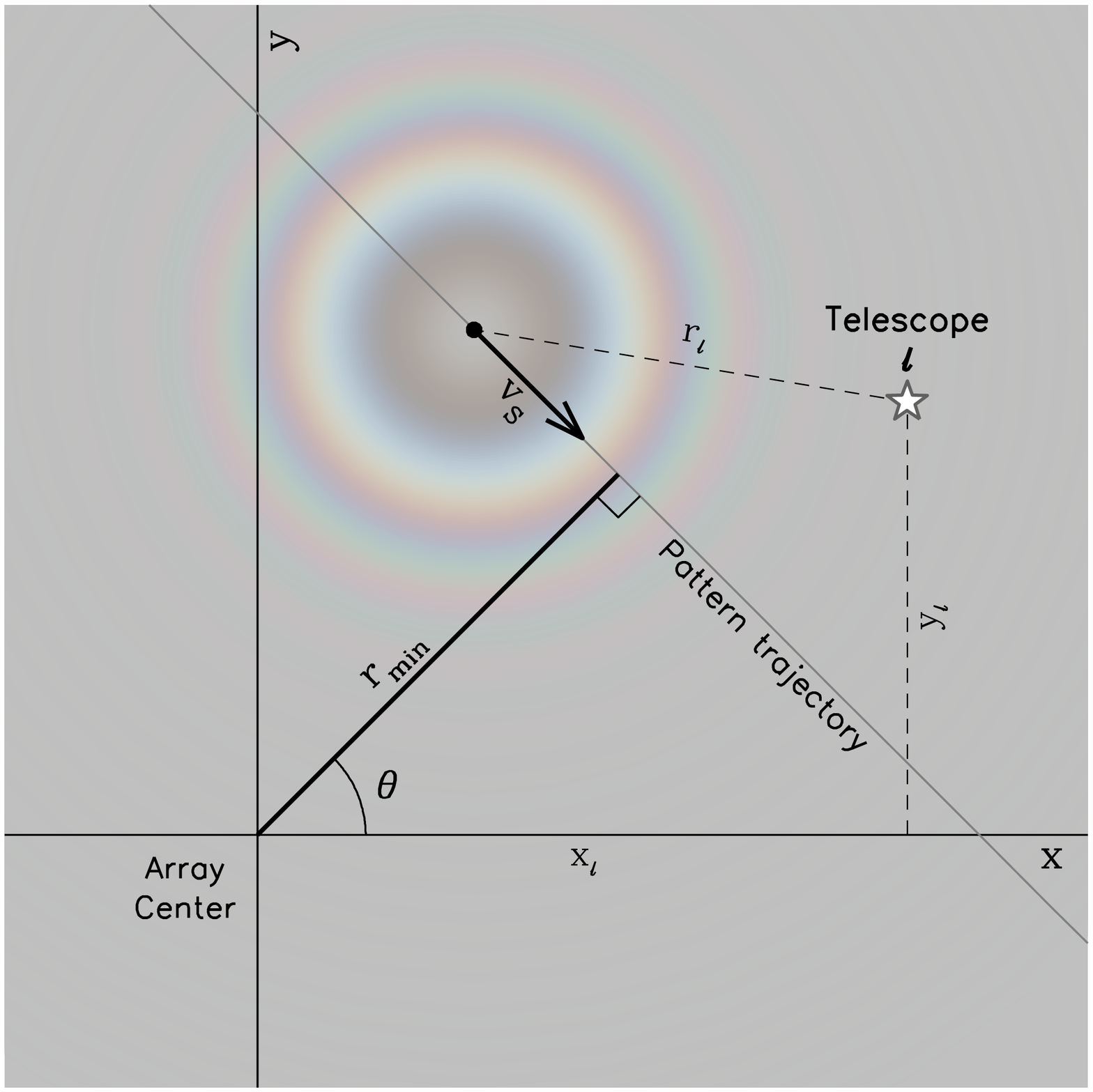}}
\caption{Geometry of the occultation event.  The locations of telescopes on Earth can be projected on to a pattern plane normal to the line of sight to the star (left).  The geometry of the event in the pattern plane is shown on right.\label{fig:ShadowGeom}}
\end{figure*}

Suppose a TJO occults a star at altitude $\mu$ and azimuth $\zeta$.  Each telescope $l$ is located at coordinates $(x_l^{\prime}, y_l^{\prime})$ on the ground, which slices through the diffraction pattern at an angle.  To calculate the intensity of the diffraction pattern at each telescope, we first project their positions on to the pattern plane that is normal to the line of sight to the TJO (Figure~\ref{fig:ShadowGeom}).  With some algebra, one can show that the coordinates of the telescope on the pattern plane is
\begin{eqnarray}
x_l & = & x_l^{\prime} / (1 + \cot^2 \mu \cos^2 \zeta)\\
y_l & = & y_l^{\prime} / (1 + \cot^2 \mu \sin^2 \zeta).
\end{eqnarray}

In the pattern plane, the diffraction pattern has a centre $(x, y)$ and moves at a speed $v$.  The pattern passes a minimum distance $r_{\rm min}$ from the origin at time $t_{\rm min}$, when it will be at angle $\theta$ with respect to the x-axis (Figure~\ref{fig:ShadowGeom}).  Thus the coordinates of the diffraction pattern in the pattern plane are 
\begin{eqnarray}
x (t) & = & r_{\rm min} \cos \theta - v \tilde{t} \sin \theta\\
y (t) & = & r_{\rm min} \sin \theta + v \tilde{t} \cos \theta,
\end{eqnarray}
where I have defined $\tilde{t} \equiv t - t_{\rm min}$.  The projected distance of the telescope $l$ from the centre of the shadow at any time is given by
\begin{eqnarray}
\nonumber r_l^2 & = & (x_l^2 + y_l^2) - 2r_{\rm min} (x_l \cos \theta + y_l \sin \theta) \\
    &   & + 2 v \tilde{t} (x_l \sin \theta - y_l \cos \theta) + r_{\rm min}^2 + v^2 \tilde{t}^2.
\end{eqnarray}
Diffraction theory gives the intensity of the diffraction pattern in terms of $r_l$.

\subsection{Measured Stellar Spectra}
The actual diffraction pattern on Earth can be considered as the sum of diffraction patterns from each point on the background star and at each wavelength.  It is this integrated diffraction pattern that photodetectors actually measure.  At {\Blue least} two properties of the background star affect the integrated diffraction pattern: the angular size of the star and {\Blue its} spectrum.

I start by using the Pickles UVILIB library of stellar spectra, which covers the entire spectral range detected by the photomultiplier tubes (PMTs) used on Cherenkov telescopes for a variety of spectral types and classes \citep{Pickles98}.  \citet{Pickles98} also gives the absolute bolometric magnitude, bolometric correction, and the effective temperature; from these quantities, I get the radius of the star.

\begin{figure}
\centerline{\includegraphics[width=8.5cm]{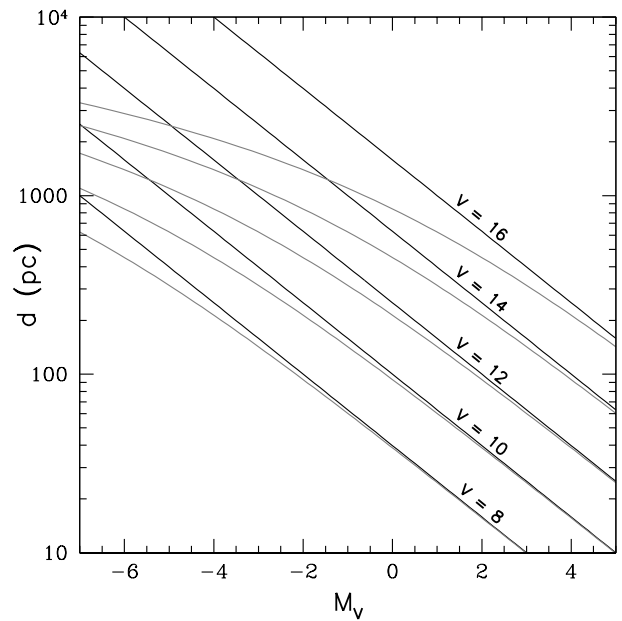}}
\caption{The distance to stars of apparent magnitudes V from 8 to 16 with $\xi = 0$ (black; no dust extinction) and $\xi = 1$ (grey).  \label{fig:ExtinctedDistances}}
\end{figure}

To convert radius into angular diameter and V-band absolute magnitude to V-band magnitude, I also need the distance to star at a given V-band magnitude.  In practice this is complicated by the presence of dust extinction, particularly on lines of sight through the Galaxy.  This effect is illustrated in Figure~\ref{fig:ExtinctedDistances}: at a given $V$, dust extinction implies that stars are nearer than expected.  The effects of dust extinction cannot be ignored for intrinsically bright stars: these are typically viewed from farther away through a larger column of dust.  To account for dust extinction, I use the extinction curve of \citet{Draine03}.  I assume that the mean density of the Galaxy is $1\xi \ \cm^{-3}$ when calculating the gas column, where $\xi$ is some scale factor.  Then, 
\begin{equation}
A_V = 0.00163\, \xi \left(\frac{D_{\star}}{\pc}\right)
\end{equation}
and the distance to the star is calculated by solving the equation
\begin{equation}
V = M_{\rm bol} - {\rm BC} + 5 \log_{10} \left(\frac{D_{\star}}{\pc}\right) + A_V.
\end{equation}

Since dust-extincted stars at a given apparent $V$ are closer to Earth than unextincted stars, they appear bigger on the sky.  This blurs out the diffraction pattern more.  I list my calculated stellar sizes for $\xi = 1$ in Table~\ref{table:StellarAngRadii}.

\begin{table*}
\begin{minipage}{170mm}
\caption{Stellar angular radii at extinction scale $\xi = 1.0$}
\label{table:StellarAngRadii}
\begin{tabular}{lcccccccccccc}
\hline
Type & \multicolumn{3}{c}{$V = 8$} & \multicolumn{3}{c}{$V = 10$} & \multicolumn{3}{c}{$V = 12$} & \multicolumn{3}{c}{$V = 14$}\\
& $D_{\star}$ & $\Theta_{\star}$ & $D_{\rm max}$ & $D_{\star}$ & $\Theta_{\star}$ & $D_{\rm max}$ & $D_{\star}$ & $\Theta_{\star}$ & $D_{\rm max}$ & $D_{\star}$ & $\Theta_{\star}$ & $D_{\rm max}$ \\ & ($\pc$) & ($\muas$) & ($\AU$) & ($\pc$) & ($\muas$) & ($\AU$) & ($\pc$) & ($\muas$) & ($\AU$) & ($\pc$) & ($\muas$) & ($\AU$)\\
\hline
O5V   & 1500 & 36  & 60  & 2200 & 24 & 130 & 3000 & 18  & 250 & 3800 & 14  & 420\\
B0V   & 970  & 27  & 110 & 1600 & 17 & 270 & 2300 & 12  & 590 & 3100 & 8.5 & 1100\\
B5V   & 310  & 35  & 64  & 630  & 17 & 250 & 1100 & 9.9 & 790 & 1700 & 6.3 & 1900\\
A0V   & 250  & 52  & 29  & 520  & 25 & 120 & {\bf 940}  & {\bf 14}  & {\bf 400} & 1500 & 8.6 & 1000\\
F0V   & 94   & 71  & 16  & 220  & 31 & 82  & 450  & 15  & 360 & 850  & 7.8 & 1300\\
G0V   & 54   & 110 & 6.6 & 130  & 46 & 37  & 290  & 20  & 180 & 580  & 10  & 750\\
K0V   & 30   & 140 & 38  & 73   & 59 & 23  & 170  & 25  & 120 & 370  & 12  & 580\\
\hline
O8III & 1600 & 48  & 34  & 2300 & 33 & 73  & 3100 & 24  & 130 & 4000 & 19  & 220\\
A0III & 330  & 55  & 26  & 650  & 28 & 100 & 1100 & 16  & 300 & 1800 & 10  & 740\\
F0III & 210  & 69  & 16  & 440  & 33 & 73  & 830  & 17  & 260 & 1400 & 10  & 710\\
G0III & 130  & 130 & 4.7 & 300  & 58 & 23  & 600  & 29  & 92  & 1100 & 16  & 290\\
K0III & 79   & 180 & 2.5 & 180  & 77 & 13  & 390  & 36  & 61  & 750  & 19  & 220\\
M0III & 400  & 570 & 0.24 & 760  & 300 & 0.88 & 1300 & 180 & 2.5 & 2000 & 120 & 5.8\\
\hline
B0I   & 2100 & 58  & 23  & 2900 & 44  & 42  & 3800 & 32  & 75  & 4700 & 26 & 120\\
A0I   & 1900 & 170 & 2.8 & 2700 & 110 & 5.5 & 3500 & 90  & 9.6 & 4500 & 71 & 15\\
F0I   & 2000 & 250 & 1.3 & 2800 & 180 & 2.5 & 3600 & 130 & 4.3 & 4500 & 110 & 6.8\\
G0I   & 1900 & 490 & 0.33 & 2700 & 350 & 0.64 & 3600 & 260 & 1.1 & 4500 & 210 & 1.7\\
K2I   & 1800 & 880 & 0.10 & 2600 & 620 & 0.20 & 3400 & 470 & 0.36 & 4300 & 370 & 0.57\\
M2I   & 2000 & 2700 & 0.011 & 2800 & 1900 & 0.021 & 3600 & 1500 & 0.037 & 4600 & 1200 & 0.058\\
\hline
\end{tabular}
\\
{\Blue $D_{\rm max}$ is the distance at which $\rho_{\star} = 1$.  Beyond this distance, the size of the stellar disc reduces the variability from the occultation.  The fiducial star is marked in bold.}
\end{minipage}
\end{table*}

Not all of the photons that reach Earth are measured by the detector.  Some are absorbed by the optics, and the detector responds to the remaining photons with some efficiency that varies with wavelength.  I assume the detection efficiency of a PMT is
\begin{equation}
\eta_{\rm detect} = \eta_{\rm optics} \times \left\{ \begin{array}{ll} 0.3 & \\
								   \multicolumn{2}{r}{(3000\ \Ang \le \lambda \le 4500\ \Ang)}\\ \\
                                                                   0.3 \times \left[4 - \displaystyle \left(\frac{\lambda}{1500\ \Ang}\right)\right] & \\
								   \multicolumn{2}{r}{(4500\ \Ang \le \lambda \le 6000\ \Ang)} \end{array} \right.
\end{equation}
and 0 at other wavelengths (based on the PMT sensitivity curve given in \citealt{Preu02}) where $\eta_{\rm optics} = 0.8$ is the throughput of the optical system of the IACTs.  

I also consider an ``ideal'' detector efficiency $\eta_{\rm detect} = 1.0$ between $3000\ \Ang$ and $6000\ \Ang$.

The measured spectrum of the star is then
\begin{equation}
\label{eqn:StarSpectrum}
F_{\lambda} = F_{\lambda}^{\rm Pickles} \times \exp(-\tau_{\rm dust}) \times \eta_{\rm detect},
\end{equation}
where $\tau_{\rm dust}$ is the optical depth of the dust as calculated using the \citet{Draine03} extinction curve.

\subsection{The Diffraction Pattern}
The ratio of a monochromatic point source's flux as the occulter passes to its unobscured flux is given by
\begin{align}
\nonumber I_{\lambda} (r_0, \rho) & = 1 + U_2^2 (\rho, r_0) + U_1^2 (\rho, r_0) + 2 \Bigl[ U_2 (\rho, r_0) \Bigr. \\
                                  & \left. \cos \left(\frac{\pi}{2} (r_0^2 + \rho^2)\right) - U_1 (\rho, r_0) \sin \left(\frac{\pi}{2} (r_0^2 + \rho^2)\right)\right]
\end{align}
when $r_0 \ge \rho$, and
\begin{equation}
I_{\lambda} (r_0, \rho) = U_0^2 (r_0, \rho) + U_1^2 (r_0, \rho)
\end{equation}
when $r_0 \le \rho$ \citep{Roques87}.  Here, $U_n$ is the $n$-th Lommel function, $r_0 = r / \ell_F$ and $\rho = R_{\rm TJO} / \ell_F$.  

Suppose the star has an angular radius $\Theta_{\star}$, with a flux $F_{\lambda}$.  The {\Blue photon flux ratio} observed by the telescope is \citep{Roques00}
\begin{align}
\nonumber I_{\star} & = \frac{2}{\pi {\Blue \Phi_{\star}} \rho_{\star}^2} \int_0^{\rho_{\star}} r^{\prime} \int_0^{\pi} \int_{\lambda_{\rm min}}^{\lambda_{\rm max}} \frac{d\Phi}{d\lambda} \\
                    &   \times  I_{\lambda} \left[\sqrt{(\lambda_0 / \lambda) (r_0^2 + r^{\prime 2} + 2 r_0 r^{\prime} \cos \psi)}, \rho (\lambda)\right] d\lambda d\psi ds_0 .
\label{eqn:FStarOriginalForm}
\end{align}
In this equation, $\rho_{\star} = \Theta_{\star} D_{\rm TJO} / \ell_F (\lambda_0)$, $r_0 = r / \ell_F(\lambda_0)$, and $d
\Phi / d\lambda\ {\Blue \equiv F_{\lambda} / (hc / \lambda)}$ is the photon number flux spectrum of the star (since the detector counts photons, not energy).  {\Blue $I_{\star}$ is normalized by the typical photon flux from the star:}
\begin{equation}
{\Blue \Phi_{\star} = \int_{\lambda_{\rm min}}^{\lambda_{\rm max}} \frac{d\Phi}{d\lambda} d\lambda.}
\end{equation}
I assume that the stellar spectrum is constant over the star's surface, ignoring limb darkening.  Then it is easiest to integrate over $\lambda$ first, getting the diffraction pattern for a TJO transiting a point source:
\begin{align}
I_{\rm point} (r_0) & = \frac{\Blue 1}{\Blue \Phi_{\star}} \int_{\lambda_{\rm min}}^{\lambda_{\rm max}} I_{\lambda} (r_0 \sqrt{\lambda_0 / \lambda}) \frac{d\Phi}{d\lambda} d\lambda,
\end{align}
and then integrating over the star's surface, which can be done efficiently as
\begin{align}
\label{eqn:IStarIntegral}
I_{\star} (r_0) & =  \frac{2}{\pi \rho_{\star}^2} \int_{|\rho_{\star} - r_0|}^{\rho_{\star} + r_0} r^{\prime} I_{\rm point} (r^{\prime}) \cos^{-1} \left[\frac{r^{\prime 2} - \rho_{\star}^2 + r_0^2}{2 r_0 r^{\prime}}\right] dr^{\prime}
\end{align}
Numerical details are given in Appendix~\ref{sec:Precision}.

\subsection{Expected noise}
\label{sec:Noise}
The two sources of noise are Poisson noise and scintillation.   The Poisson noise in the number of photons is simply $\sigma_n^{\rm Poisson} = \sqrt{n_{\rm sky} + n_{\star}}$.  The number of photons $n_{\star}$ from the star in a time bin is calculated from the stellar spectra using equation~\ref{eqn:StarSpectrum}.  To calculate the number of photons $n_{\rm sky}$ from the sky{\Blue,} I use the sky spectrum measured at Kitt Peak in \citep{Neugent10}.  I extend this spectrum to wavelengths below 3700 \AA~by assuming the sky has an AB magnitude of 22.5 per square arcsecond in this range.  The spectrum is then normalized to $2 \times 10^{12}$ photons per square meter per steradian {\Blue per second} between 3000 \AA~and 6500 \AA~\citep{Deil09}.  The number of sky photons is then calculated by multiplying by the solid angle of the sky covered by one PMT pixel (Table~\ref{table:IACTPixels}) after convolving with the PMT sensitivity (as done for the star in equation~\ref{eqn:StarSpectrum}).

\begin{table}
\begin{minipage}{90mm}
\caption{IACT Pixel Sizes}
\label{table:IACTPixels}
\begin{tabular}{lcccccc}
\hline
Array   & Telescope Aperture & Pixel diameter & References\\
        & $m$                & $^{\circ}$    & \\
\hline
VERITAS & 12.0 & 0.15  & (1)\\
HESS    & 12.0 & 0.16  & (2)\\
        & 28.0 & 0.067 & (3)\\
CTA     & 7.2  & 0.25  & (4)\\
        & 12.3 & 0.18  & (4)\\
	& 24.0 & 0.09  & (4)\\
\hline
\end{tabular}
\\{\bf References} -- (1) \citet{Weekes02}; (2) -- \citep{Bernlohr03}; (3) http://www.mpi-hd.mpg.de/hfm/HESS; (4) \citet{Bernlohr13}
\\
\end{minipage}
\end{table}

In terms of magnitudes, the scintillation noise is
\begin{align}
\nonumber \sigma_{\rm mag}^{\rm scint} & = 5.6 \times 10^{-4} X^{7/4} \left(\frac{A_{\rm tel}}{100\ \meter^2}\right)^{-1/3} \left(\frac{\Delta t}{1\ \sec}\right)^{-1/2}\\
& \times \exp\left(-\frac{h}{8\ \km}\right),
\end{align}
where $A_{\rm tel}$ is the collecting area of the telescope, $X$ is the airmass, and $h$ is the altitude of the telescope \citep{Young67,Southworth09}.  This can be converted into the noise in the number of photons by noting that $\sigma_{\rm mag} \approx 1.0857 \sigma_n$, implying $\sigma_n^{\rm scint} = \sigma_{\rm mag}^{\rm scint} n_{\rm star} / 1.0857$.  Then the total noise is
\begin{equation}
\label{eqn:sigman}
\sigma_n^2 = (\sigma_n^{\rm Poisson})^2 + (\sigma_n^{\rm scint})^2.
\end{equation}

{\Blue As forms of white noise, both Poisson noise and scintillation noise fall as integration time increases, but any pink or red noise correlated between time samples adds additional uncertainties \citep*{Pont06}.  Scintillation, for example, is known to have a pink noise component \citep*{Bickerton09}.  In addition, there can be variability on time-scales much larger than a fraction of a second.  The geometry of stray light changes during the course of a night and probably spoils photometry on hour time-scales (Deil, private communication via \citealt{Lacki11}).}

{\Blue The $\gg \msec$ stability of stellar flux levels in Cherenkov telescopes is not reported often, but \citet{Deil09} does describe the power spectral density of the variability for frequencies $0.1$--$1000\ \Hz$ as observed in HESS with a specially built optical camera.  They find that the noise at high frequencies ($\ga 100\ \Hz$) is indeed white, but a $1/f$ noise component dominates at lower frequencies.  They attribute the pink noise to electronic noise and note that it is dependent on temperature.}

{\Blue Pink noise, with a $1/f$ power spectrum, has equal power per log bin of time \citep{Schroeder91}.  The approximate magnitude variability of stars at time-scales $0.01$ -- $10\ \sec$ is then roughly the same as the Poisson/scintillation noise at $100\ \Hz$ frequencies.  The expected 100 Hz variability is $\sim 3\%$, while the occultation causes fluctuations of up to $\sim 20\%$ in the photon count rate, suggesting that the occultations should be detectable.  For comparison, \citet{Hinton06} present the light curves of a meteor flash in the HESS optical curves which display little variability on $0.1\ \sec$ scales.}

{\Blue All of this assumes that the noise properties in the normal Cherenkov PMTs (or other detectors) are similar as in the HESS optical camera system, though.  The PMTs used in most of the Cherenkov telescope pixels have different setups, with significant deadtime, for example (Benbow, private communication).  The power spectrum of the count rate variability in IACTs should be investigated at $\sim 10\ \Hz$ to be sure.}

\subsection{The fiducial model}
Idealized as it is, there are many parameters in these models, including the stellar type, magnitude, extinction, and the TJO's distance and size.  To get a sense of how the detectability and parameter estimation varies with these parameters, I consider a ``fiducial'' model and then vary only one or two of the free parameters at a time from that base.

This particular model is supposed to represent a typical occultation by a Kuiper Belt object with reasonable signal-to-noise.  The fiducial object is located $10^{1.6}$ AU away and has a radius of $10^{2.5}$ m.  This is an object slightly closer and slightly smaller than those of the two TJO occultations detected by HST \citep{Schlichting09,Schlichting12}.  The fiducial star has a spectral type of A0V, a V-band magnitude of 12, and suffers extinction with $\xi = 1$.  This is the most commonly used star type in {\Blue N07}, except with extinction.  In addition, the confusion limit for IACTs is $V \approx 14$ \citep{Lacki11}; $V = 12$ stars should be individually detectable by the IACT.  The star is assumed to be at an altitude $\mu = 90^{\circ}$.  I also assume the diffraction pattern passes through the centre of the array with $r_{\rm min} = 0$ at a speed $30\ \kms$ {\Blue in the pattern plane}. 

\begin{figure*}
\centerline{\includegraphics[width=8.5cm]{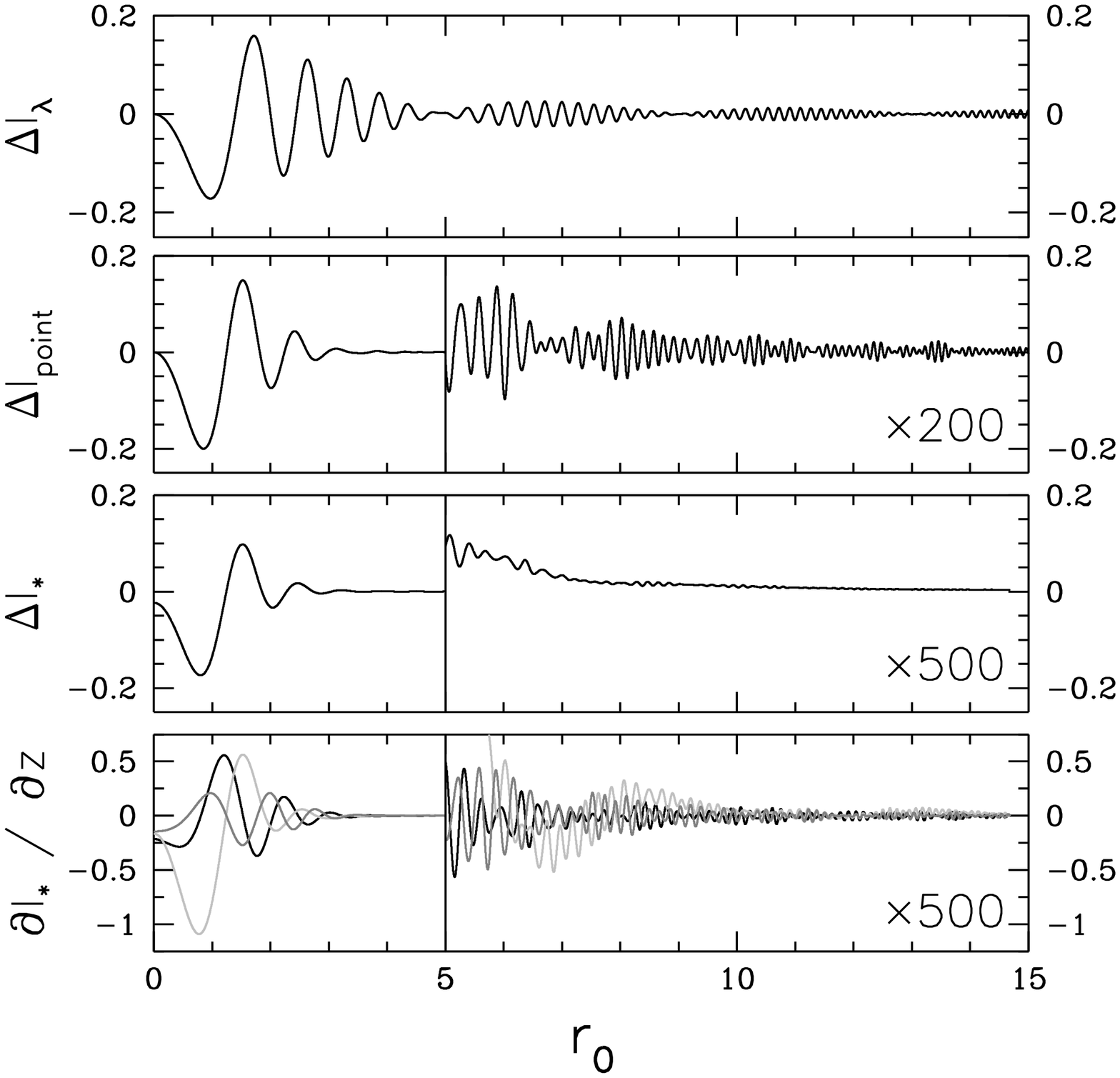}\ \includegraphics[width=8.5cm]{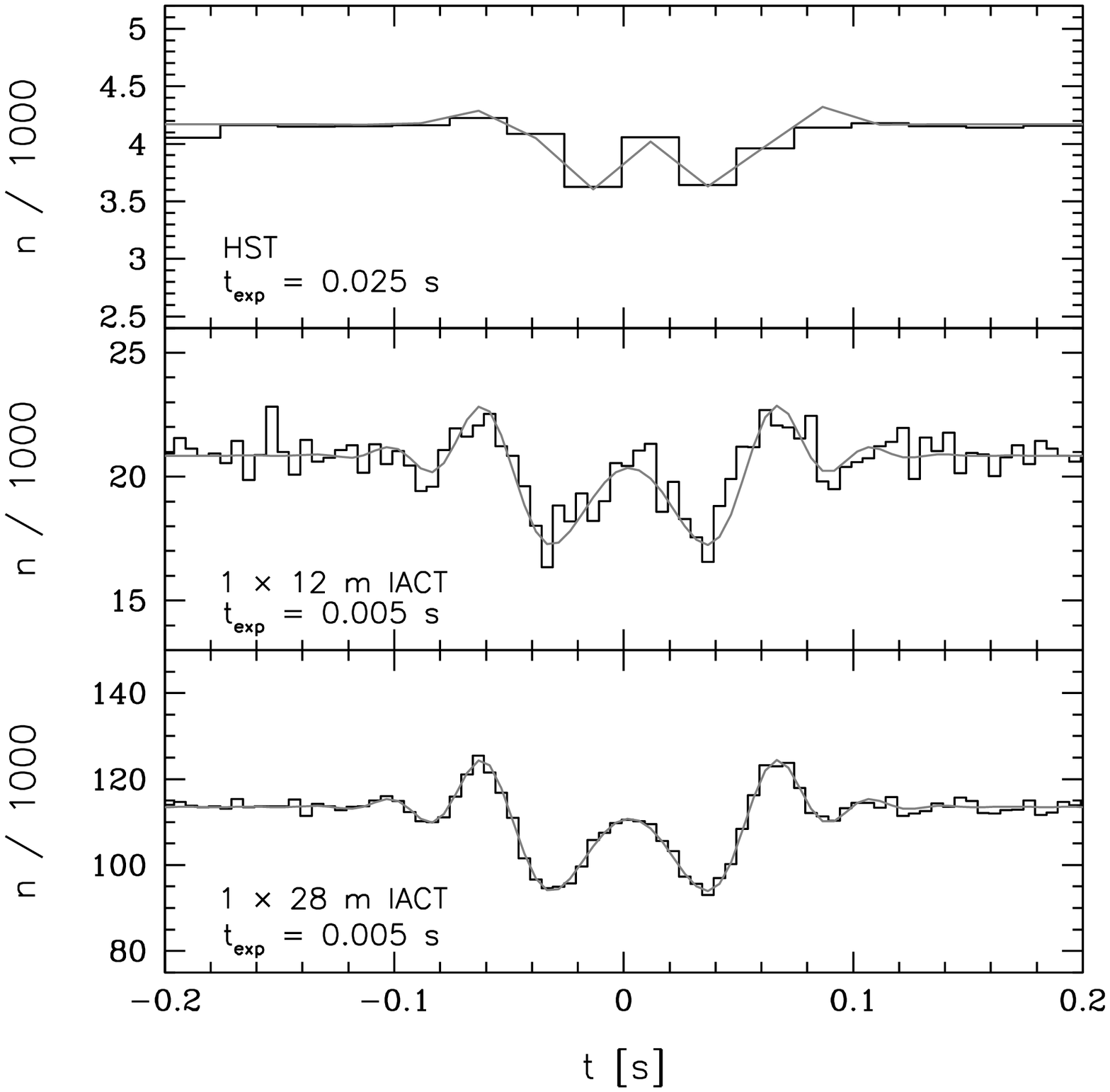}}
\caption{On left, the theoretical light curves of the fiducial occultation and the partial derivatives of $I_{\star}$ with respect to $r_0$ (black), $\rho$ (dark grey), and $\rho_{\star}$ (light grey).  The values of $\Delta I_{\rm point}$, $\Delta I_{\star}$, and the $\partial I_{\star} / \partial \{r_0, \rho, \rho_{\star}\}$ are exaggerated for $r_0 \ge 5$ to show the detail. On right, simulated light curves of the fiducial occultation for various instruments.\label{fig:FiducialModel}}
\end{figure*}

I show the expected light curves for the fiducial {\Blue event} in Figure~\ref{fig:FiducialModel}.  Note that the ringing in the monochromatic point-source diffraction pattern $I_{\lambda}$ is suppressed by integration over wavelength and then by integration over the star's disc.  I also show a realization of the light curve with random noise (both Poisson and scintillation, treated as Gaussian white noise; Section~\ref{sec:Noise}) added as viewed by HST (top), a single 12 meter IACT as used in VERITAS (middle), and a single 28 meter IACT as used by HESS (bottom).  The diffraction fringes are clearly detected by the IACTs.  Note that I use a finer time resolution for the IACTs (200 Hz) than is possible with Hubble (40 Hz).  Also note that IACTs have multiple telescopes, enhancing the signal further.  

For comparison purposes, I also consider a ``bright'' star with $V = 10$, spectral type B5V, and $\xi = 0$.  

\section{Which Events are Detectable?}
\label{sec:Detectability}

\subsection{Detectability Using Matched Filters}
\label{sec:MatchedFilter}
{\Blue N07 included the most extensive discussion of the detectability of TJO occultations.  They defined a detectability statistic $\Xi$ that is basically the deviation of $I_{\star}$ from 1 integrated over time in units of a Fresnel-crossing time.  Then if the event occurs over $m$ time bins, $\chi^2 = (n_k - \bar{n}_{\star})^2 / \sigma^2$ has a mean $\bar{n}_{\star} \Xi + m$, after defining $n_k$ as the number of photons in time bin $k$ and $\bar{n}_{\star}$ as the expected number of photons when there is no event.  The chance that the variations in the photon count rate are as large as the observed in the signal can then be estimated by noting that $\chi^2$ has a mean and variance of $m$ if no occultation occurs.}

{\Blue The N07 statistic provides a robust and conservative method of detecting any abnormally large fluctuation in the count rate, but it is not optimized for TJO occultations specifically.  Occultations produce a characteristic ringing structure in the light curve, with $I_{\star}$ changing smoothly from moment to moment.  In contrast, the vast majority of signals that the $\chi^2$ statistic can detect are essentially noise with abnormally big variance and no time structure at all.  The $\chi^2$ method loses statistical power by searching for a vast set of possible signals, of which occultations events are a tiny subset.  This is reflected in the dependence of $\chi^2$ on $m$; the statistic actually decreases as the sampling rates passes $v / \ell_F$ due to the greater number of possible signals.}

We optimize the signal-to-noise ratio for a {\Blue \emph{known}} signal, in the presence of additive Gaussian white noise,\footnote{For a data series, the noise is white if the noise in each measurement is statistically independent.} by applying a matched filter to the data \citep[e.g.,][]{Davis89}.  Consider a series of data $x(t) = s(t) + w(t)$, where $s(t)$ is an intrinsic signal we are interested in measuring, $w(t)$ is additive white noise with constant variance, and $x(t)$ averages to 0.  If we wish to detect a known template pattern $\tau(t)$, we cross-correlate the observed data with a matched filter, which is simply $\tau(t)$ itself: ${\cal C} = \int x(t) \tau (t) dt$.  The correlation ${\cal C}$ is largest when the intrinsic signal $s(t)$ matches the template signal $\tau(t)$.  For an occultation, the data and the templates are the light curve of the background star (this is the cross-correlation method of \citealt*{Bickerton08}).  Note that both the data and template must be normalized so that the noise is constant (that is, the noise must be whitened).  

Suppose we wish to detect a fluctuation $\Delta I_{\rm template}$ in the star's light curve.  The renormalized data are 
\begin{equation}
\hat{n}_{\star} (t_k; l) = \Delta I_{\rm obs} (t_k; l) \frac{\bar{n}_{\star} (l)}{\sigma_{n} (l)},
\end{equation}
where $\bar{n}_{\star} (l)$ is defined for telescope $l$, $\Delta I_{\rm obs} (t_k; l)$ is the ratio of the observed brightness of the star and its baseline brightness, and $\sigma_{n} (l)$ is the noise in the photon rate.  For simplicity, I ignore variation in the noise rate during the event; this is valid for occultations with small depths.  We cross-correlate with a template 
\begin{equation}
\hat{\tau}_l (t_k; l) = \Delta I_{\rm template} (t_k; l) \frac{\bar{n}_{\star} (l)}{\sigma_{n} (l)}
\end{equation}
to get the quantity:
\begin{equation}
\nu = \sum_l \sum_k \Delta I_{\rm template} (t_k; l) \Delta I_{\rm obs} (t_k; l) \left(\frac{\bar{n}_{\star} (l)}{\sigma_{n} (l)}\right)^2 dt.
\end{equation}
Note that if there is no event ($\Delta I_{\rm obs} (t_k; l) = 0$), the expectation value for $\nu$ is 0. 
 
As long as the noise in each time bin is uncorrelated and has a normal distribution, $\nu$ itself is normally distributed.  {\Blue The} expected variance of $\nu$ {\Blue is}
\begin{align}
\sigma_{\nu}^2 & = \sum_l \sum_k \left[\Delta I_{\rm template} (t_k; l) \left(\frac{\bar{n}_{\star} (l)}{\sigma_{n} (l)}\right)\right]^2.
\end{align}
The effective signal-to-noise ratio (SNR) is then just $\nu / \sigma_{\nu}$.  The signal is detected with significance level $p$ if
\begin{equation}
p \ge \frac{1}{2} {\rm erfc} \left(\frac{\nu}{\sqrt{2}\sigma_{\nu}}\right).
\end{equation}

The cost of the matched filter approach is that the actual parameters of the occultation patterns are unknown, so a large number of templates must be tried. Therefore, we must set $p$ small to account for this look-elsewhere effect: essentially we must set $p$ to the reciprocal of the number of observations times the number of model occultation patterns.   In order for the matched filter to work properly, the ringing of the template signal must be in phase with the ringing of the observed signal.  The ratio of observing time to the duration of a single occultation event is $\sim 10^{10}$ \citep{Lehner09}.  In addition, there are a number of parameters that must be fit: $r_{\rm min}$, $\theta$, $v$, {\Blue $\rho$,} and $\rho_{\star}$.  If there are $\sim$100 choices for each parameter, then there are $\sim 10^{10}$ possible templates.  I thus set a threshold of $p = 10^{-20}$, which requires $\nu/\sigma_{\nu} \ge 9.26$.

\subsection{Results}
\begin{figure*}
\centerline{\includegraphics[width=8.5cm]{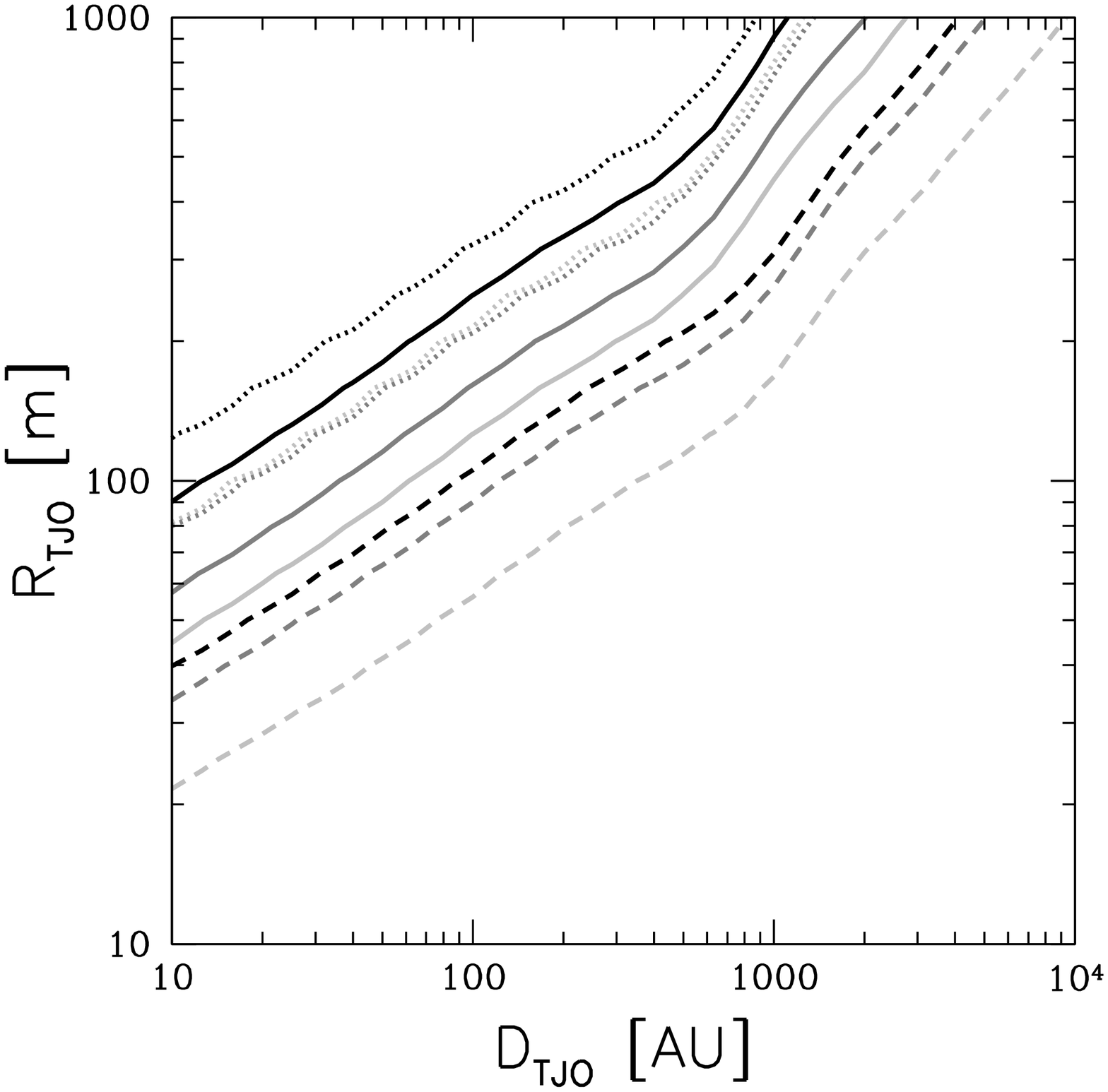}\ \includegraphics[width=8.5cm]{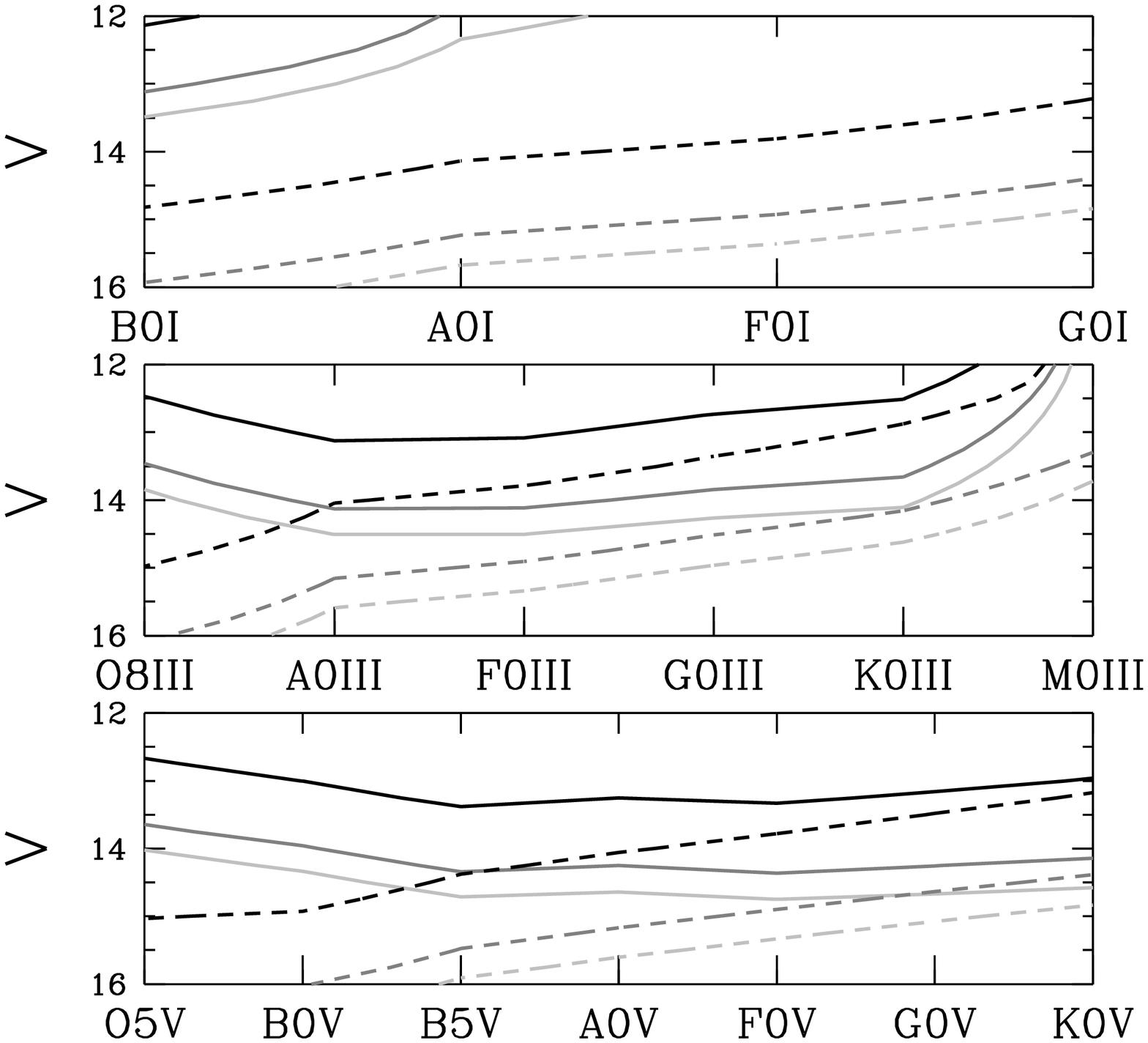}}
\caption{{\Blue On left,} the sensitivity of IACT arrays to occultations by TJOs for a {\Blue speed $v = 30\ \kms$.  VERITAS is black, HESS is dark grey, and CTA is light grey.  The solid lines are for the matched statistic with $p \le 10^{-20}$ if the target star is $V = 12$ A0V with $\xi = 1$, whereas the dashed lines are the same statistic but for the ``bright'' (B5V, $V = 10$, $\xi = 0$) star.  For comparison, the dotted lines are using the N07 $\chi^2$ statistic with $p \le 10^{-11}$.  On right, t}he sensitivity of IACT arrays to occultations by a $10^{2.5} \approx 300$ m radius TJO at $10^{1.6} \approx 40$ AU for different magnitude and stars with $\xi = 1$.  The line styles are the same, except that the {\Blue dashed} lines now stand for $\xi = 0$.  \label{fig:DetectabilityDR}\label{fig:DetectabilityStarV}}
\end{figure*}

IACT arrays are excellent detectors of sub-kilometre TJOs.  I show the limits on object radius and distance that can be detected with in Figure~\ref{fig:DetectabilityDR}.  For the occultations of ``bright'' B5V $V = 10$ stars, VERITAS is able to detect objects as small as $80\ \meter$ in radius 40 AU away, and 1 km radius objects out to 4000 AU (0.02 pc).  Even for the less ideal A0V $V = 12$ star, VERITAS can detect 200 meter radius objects 40 AU away and 1 km radius objects at 1000 AU (0.004 pc).  But these limits pale compared to those for HESS, which has a large central telescope that can collect many photons and has relatively low scintillation noise.  For the fiducial (bright) star, HESS can detect objects with $R_{\rm TJO} \ga 130\ \meter$ ($60\ \meter$) that are 40 AU away, and 1 km radius objects with $D_{\rm TJO} \la 2000\ \AU$ ($\la 5000\ \AU$).  The reach of CTA will be even more phenomenal.  It is sensitive to occultations of the fiducial (``bright'') star by $R_{\rm TJO} \ga 100\ \meter$ ($\ga 40\ \meter$) objects 40 AU away and 1 km radius objects that are $D_{\rm TJO} \la 2500\ \AU$ ($\la 8000\ \AU$) away.

To put the sensitivities in perspective, what are the sensitivities of these arrays to the two TJO occultations observed with \emph{Hubble}?  These occultations, by $R_{\rm TJO} \approx 500\ \meter$ objects roughly $D_{\rm TJO} \approx 45\ \AU$ away \citep{Schlichting09,Schlichting12}, would have had SNRs of 63 (320) at VERITAS, 150 (430) at HESS, and 240 (1100) at CTA for the fiducial (``bright'') star. Likewise, the fiducial object has a significance of 31 (160) at VERITAS, 74 (220) at HESS, and 120 (550) at CTA when it occults the fiducial (``bright'') star.

Increasing the photon detection efficiency and decreasing the pixel size of the IACT detectors leads to strong improvements in their sensitivity to occultations of fainter stars.  {\Blue The} use of ``ideal'' detectors {\Blue improves} the sensitivity of the IACT arrays to occultations of the fiducial star to those of the ``bright'' star.  However, the ``ideal'' detectors do not improve sensitivity to occultations of the ``bright'' stars, because then the noise is dominated by scintillation.

So far, I have assumed the background star is A0V or B5V -- as relatively nearby and blue star types, these stars have small angular diameters and therefore are well-suited for occultation detection (Table~\ref{table:StellarAngRadii}).  The sensitivity {\Blue to} occultations of other star types is shown in Figure~\ref{fig:DetectabilityStarV}.  As it turns out, all B to K dwarfs are roughly equally suitable for occultations: VERITAS is sensitive to occultations of $V = 12.5$ dwarfs and CTA is sensitive to $V = 14$ dwarfs by the fiducial object.  In fact, O5V stars are actually \emph{worse} than K dwarfs, because they are so strongly affected by dust extinction (Figure~\ref{fig:ExtinctedDistances}).  Moving to the brighter giant stars, I again find that occultations of A to K dwarfs are roughly equally detectable, with slightly poorer sensitivity for B giants, and low sensitivity for M giants.  The magnitude limits remain similar to dwarfs.  But the sensitivities are especially poor for supergiants, where the sensitivity drops by several magnitudes.  The best sensitivity for occultations of supergiants are when the supergiants are early type.

\begin{figure}
\centerline{\includegraphics[width=8.5cm]{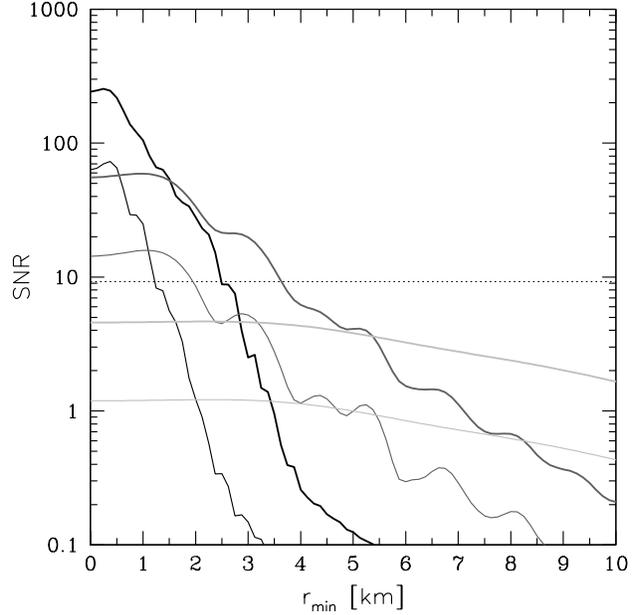}}
\caption{The sensitivity of IACT arrays to occultations by the fiducial object as $r_{\rm min}$ increases.  The thick lines are for CTA and the thin lines are for VERITAS.  The three cases shown are $D_{\Blue \rm TJO} = 10\ \AU$ (black), $100\ \AU$ (dark grey), and $1000\ \AU$ (light grey).\label{fig:DetectabilityrMin}}
\end{figure}

The diffraction pattern of an occultation need not pass through the core of an IACT array; it may pass some distance away, with a larger $r_{\rm min}$, so that only the outer fringes are detected by the IACT array.  Yet the IACT array can still detect the occultation from the outer fringes alone.  In Figure~\ref{fig:DetectabilityrMin}, I show how the detection significance varies with $r_{\rm min}$ for the fiducial occultation.  The significance is essentially constant as long as $r_{\rm min} \la l_F$.  At larger distances, the significance falls off exponentially as only the outer fringes are detected.  But even then the IACTs can still detect diffraction patterns that pass several kilometres away, depending on the depth of the event, simply because the light collecting power of the IACTs is so great.

Intuitively, $\nu$ increases linearly with the number of photons collected during an event.  Therefore, it scales inversely with the projected speed of the diffraction pattern.  This is confirmed by an actual calculation.

{\Blue For comparison, the limits of detectability using the N07 $\chi^2$ statistic are plotted as the dotted lines in Figure~\ref{fig:DetectabilityDR}.  I choose a time bin of $0.1 \ell_F / v$, and I set the $p$ threshold at $10^{-11}$ (N07).  As expected, the power of the IACTs is weakened using this statistic.  In fact, CTA is no more powerful than HESS, since it uses more telescopes, meaning more observations and more degrees of freedom.  The fiducial event is still detectable even with VERITAS, so the ability of Cherenkov telescopes to find Kuiper Belt objects of this statistic seems robust.}

\section{Parameter Estimation with IACTs}
\label{sec:ParameterEstimation}

\subsection{Fisher matrices}
\label{sec:FisherMatrices}
The expected covariance matrix for estimated parameters of an experiment can be predicted beforehand by calculating its Fisher matrix \citep[for a previous example of Fisher matrices applied to TJO occultations, see][]{Cooray03-Fisher}.  Suppose an experiment makes $K$ measurements $z_k$, with the $k$-th measurement having an rms error of $\sigma_k$.  It is fit by a model with $M$ parameters $\alpha_m$.  Given the particle derivatives of the expected observed values with each parameter, the Fisher matrix is constructed as:
\begin{equation}
{\cal F}_{ij} = \sum_{k = 0}^K \frac{1}{\sigma_k^2} \frac{\partial z_k}{\partial \alpha_i} \frac{\partial z_k}{\partial \alpha_j}
\end{equation}
where $i$ and $j$ are in the range 1 to $m$.  The expected covariance matrix is just the inverse of the Fisher matrix, with $\sigma_{ij} = ({\cal F}^{-1})_{ij}$ for parameters $\alpha_i$ and $\alpha_j$.  

To calculate the Fisher matrix, we need the partial derivative of $I_{\star}$ with each parameter.  The $I_{\star}$ function varies only if $r_0$, $\rho$, or $\rho_{\star}$ varies; therefore the partial derivatives we need can be constructed from $\partial I_{\star}/\partial r_0$, $\partial I_{\star}/\partial \rho$, and $\partial I_{\star}/\partial \rho_{\star}$.  The formulae for two of the derivatives for a monochromatic point source are given in Appendix~\ref{sec:Derivations}.  Then, again assuming that the star's spectrum is constant across its surface, we can first integrate over wavelength:
\begin{align}
\frac{\partial I_{\rm point}}{\partial r_0} (r_0) & = \int_{\lambda_{\rm min}}^{\lambda_{\rm max}} \frac{\partial I_{\lambda}}{\partial r_0} (r_0 \sqrt{\lambda_0 / \lambda}) \frac{d\Phi}{d\lambda} d\lambda\\
\frac{\partial I_{\rm point}}{\partial \rho} (r_0) & = \int_{\lambda_{\rm min}}^{\lambda_{\rm max}} \frac{\partial I_{\lambda}}{\partial \rho} (r_0 \sqrt{\lambda_0 / \lambda}) \frac{d\Phi}{d\lambda} d\lambda,
\end{align}
and then over the star's surface:
\begin{align}
\frac{\partial I_{\star}}{\partial r_0} (r_0) & =  \frac{2}{\pi \rho_{\star}^2} \int_{|\rho_{\star} - r_0|}^{\rho_{\star} + r_0} r^{\prime} \frac{\partial I_{\rm point}}{\partial r_0} (r^{\prime}) \cos^{-1} \left[\frac{r^{\prime 2} - \rho_{\star}^2 + r_0^2}{2 r_0 r^{\prime}}\right] dr^{\prime}\\
\frac{\partial I_{\star}}{\partial \rho} (r_0) & =  \frac{2}{\pi \rho_{\star}^2} \int_{|\rho_{\star} - r_0|}^{\rho_{\star} + r_0} r^{\prime} \frac{\partial I_{\rm point}}{\partial \rho} (r^{\prime}) \cos^{-1} \left[\frac{r^{\prime 2} - \rho_{\star}^2 + r_0^2}{2 r_0 r^{\prime}}\right] dr^{\prime}.
\end{align}
Finally, the derivative with respect to stellar radius is found by differentiating equation~\ref{eqn:FStarOriginalForm}:
\begin{align}
\frac{\partial I_{\star}}{\partial \rho_{\star}} (r_0) & = \frac{-2 I_{\star}}{\rho_{\star}} + \frac{2}{\pi \rho_{\star}} \int_0^{\pi} I_{\rm point} (\sqrt{r_0^2 + r^{\prime 2} + 2 r_0 r^{\prime} \cos \psi}) d\psi.
\end{align}

\subsubsection{\Blue Theorist's parameter set}
Between the observing geometry and the diffraction pattern, my model has seven unknown parameters.  {\Blue There is some freedom in choosing which seven variables stand in for these parameters.  From a theoretical perspective, we are interested in the physical properties of the occulter, such as its distance, size, and speed.  I first consider a set of ``theorist's'' parameters -- $r_{\rm min}$, $v$, $\theta$, $t_{\rm min}$, $R_{\rm TJO}$, $D_{\rm TJO}$, $\Theta_{\star}$ -- explained below.}

{\Blue   The partial derivatives are calculated by assuming that the other variables in the parameter set are kept constant.  For example, suppose we had a set of two parameters, $\{\bar{\rho} \equiv 2 \rho, \ell_F\}$.  Then $\partial I_{\star} / \partial \bar{\rho}$ is calculated assuming $\ell_F$ (and thus $r_F$) is constant.  A pure increase in $\bar{\rho}$, with $\ell_F$ fixed, means that the TJO is bigger, because decreasing the distance would affect $\ell_F$ too.  But if I used $R_{\rm TJO}$ instead of $\ell_F$, I would calculate $\partial I_{\star} / \partial \bar{\rho}$ assuming that $R_{\rm TJO}$ is constant: a pure increase in $\bar{\rho}$ for this parameter set means the TJO is closer, because changing the size would affect $R_{\rm TJO}$ as well.  It is like partial derivative in thermodynamics, where one must specify which variables are being held constant.}

The first four {\Blue variables} relate solely to the observing geometry: $r_{\rm min}$, $v$, $\theta$, and $t_{\rm min}$.  None of these depends on the Fresnel-scaled size of the occulter or the star; all can be expressed in terms of $r_0$:
\begin{align}
\frac{\partial I_{\star}}{\partial r_{\rm min}} & = \frac{\partial I_{\star}}{\partial r_0} \frac{x_l \cos \theta + y_l \sin \theta + r_{\rm min}}{r_l \ell_F}\\
\frac{\partial I_{\star}}{\partial v}           & = \frac{\partial I_{\star}}{\partial r_0} \frac{\tilde{t} (x_l \sin \theta - y_l \cos \theta) + v \tilde{t}^2}{r_l \ell_F}\\
\frac{\partial I_{\star}}{\partial \theta}      & = \frac{\partial I_{\star}}{\partial r_0} \frac{r_{\rm min}(x_l \sin \theta - y_l \cos \theta) + v \tilde{t} (x_l \cos \theta + y_l \sin \theta)}{r_l \ell_F}\\
\frac{\partial I_{\star}}{\partial t_{\rm min}} & = \frac{\partial I_{\star}}{\partial r_0} \frac{-v (x_l \sin \theta - y_l \cos \theta) - v^2 \tilde{t}}{r_l \ell_F}.
\end{align}
{\Blue These} derivatives of $I_{\star}$ follow simply from the Chain Rule.  The next two parameters to be fit are the size of the body and the angular size of the star.  The body's radius is simply $\rho$ times the Fresnel scale.  {\Blue Since $D_{\rm TJO}$ is independent of $R_{\rm TJO}$, there is no dependence on $\rho_{\star}$ or $r_0$, so that:}
\begin{align}
\frac{\partial I_{\star}}{\partial R_{\rm TJO}} = \frac{\partial I_{\star}}{\partial \rho} \frac{1}{\ell_F}.
\end{align}
Likewise, the angular size of the star solely depends on $\rho_{\star}$: $\Theta_{\star} = \rho_{\star} \ell_F / D_{\rm TJO}$.  This gives
\begin{align}
\frac{\partial I_{\star}}{\partial \Theta_{\star}} = \frac{\partial I_{\star}}{\partial \rho_{\star}} \frac{D_{\rm TJO}}{\ell_F}.
\end{align}
Finally, the distance is unknown.  All three of the variables $r_0$, $\rho$, and $\rho_{\star}$ that determine the value of $I_{\star}$ are physical quantities scaled by $\ell_F$, which in turn depend on the distance.  Therefore, by the Chain Rule, we have
\begin{align}
\nonumber \frac{\partial I_{\star}}{\partial D_{\rm TJO}} & = \frac{\partial I_{\star}}{\partial r_0} \frac{\partial r_0}{\partial D_{\rm TJO}} + \frac{\partial I_{\star}}{\partial \rho} \frac{\partial \rho}{\partial D_{\rm TJO}} + \frac{\partial I_{\star}}{\partial \rho_{\star}} \frac{\partial \rho_{\star}}{\partial D_{\rm TJO}}\\
& = \frac{1}{2 D_{\rm TJO}} \left[\rho_{\star} \frac{\partial I_{\star}}{\partial \rho_{\star}}  - r_0 \frac{\partial I_{\star}}{\partial r_0} - \rho \frac{\partial I_{\star}}{\partial \rho}\right].
\end{align}

\subsubsection{\Blue Observer's parameter set}
{\Blue The theorist's parameter set is not the best formulation from an observational point of view.  What the IACTs can actually measure is the duration of the event, the size of the Fresnel pattern, and the shape of the light curve (which depends on $\rho$ and $\rho_{\star}$).  These are essentially independent of one another, and should have small covariances.  I therefore consider a set of observer's parameters.  The first three are $r_{\rm min}$, $\theta$, and $t_{\rm min}$, as before.  Then we have $\ell_F$, $\nu_{\rm occ} = \ell_F / v$, $\rho$, and $\rho_{\star}$.  As with the theorist's variables, I calculate the partial derivatives with respect to each parameter using the Chain Rule.}

{\Blue The partial derivatives for $r_{\rm min}$, $\theta$, and $t_{\rm min}$ have the same values as for the physical parameter set.  The partial derivatives $\partial I_{\star} / \partial \rho$ and $\partial I_{\star} / \partial \rho_{\star}$ are also the same as those calculated in the beginning of the section.  The remaining variables $\nu_{\rm occ}$ and $\ell_F$ do not depend on the size of the object or the star.  They are 
\begin{align}
\frac{\partial I_{\star}}{\partial \nu_{\rm occ}} & = \frac{\partial I_{\star}}{\partial r_0} \frac{\tilde{t} (x_l \sin \theta - y_l \cos \theta + v \tilde{t})}{r_l}\\
\frac{\partial I_{\star}}{\partial \ell_F} & = -\frac{\partial I_{\star}}{\partial r_0} \left[\frac{r_l}{\ell_F} - \frac{\nu_{\rm occ} \tilde{t}(x_l \sin \theta - y_l \cos \theta + \tilde{t} v)}{r_l} \right].
\end{align}
}

\subsection{Fisher matrix results}
\label{sec:FisherResults}
\begin{figure*}
\centerline{\includegraphics[width=18cm]{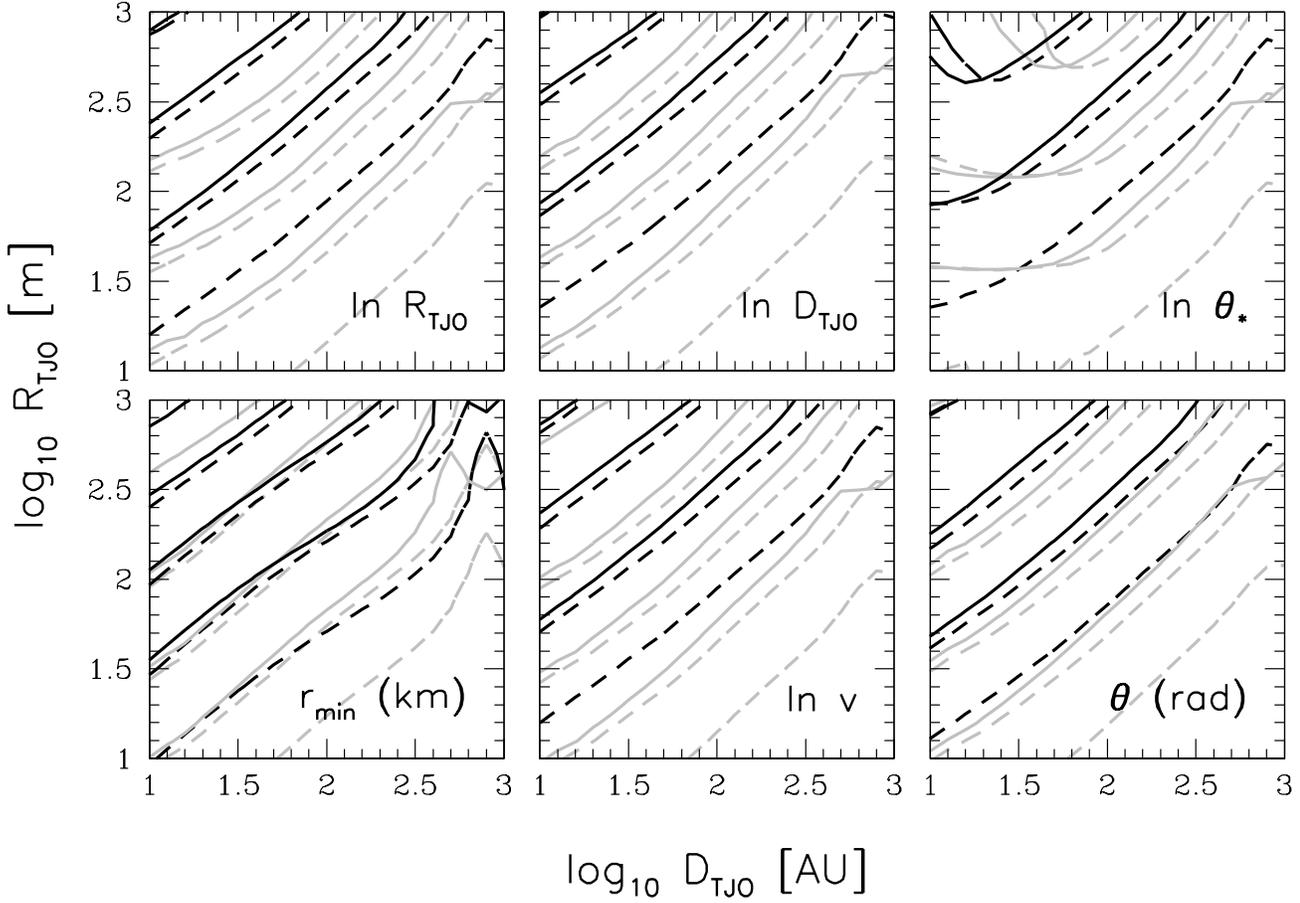}}
\caption{IACT arrays can estimate the speed, position, size, and distance of an occulting object, as well as the angular size of the background star.  These are the {\Blue projected uncertainties} for objects occulting A0V $V = 12$ $\xi = 1$ stars ({\Blue solid}) and B5V $V = 10$ $\xi = 0$ stars ({\Blue dashed}) with $v = 30\ \kms$.  {\Blue The contours are for VERITAS (black) and CTA (light grey); HESS' performance is comparable to VERITAS.}  Contours {\Blue indicate, from lower right to upper left, uncertainties of 1, 0.1, 0.01, and 0.001}.\label{fig:ParamErrorsVERITAS}\label{fig:ParamErrorsCTA}}
\end{figure*}

Although one potential advantage of IACT arrays is breaking degeneracies in parameter fitting, not every detectable event is tightly constrained.  Figure~\ref{fig:ParamErrorsVERITAS} shows the precision IACT arrays can reach for each {\Blue theorist's} parameter for TJOs of different sizes and distances.  I find that the precision drops rapidly with distance.  Note that occultations of ``bright'' stars improve parameter estimation precision by an order of magnitude (grey lines); therefore these occultations are the best chance that VERITAS and HESS have of measuring the physical properties of TJOs.  {\Blue Roughly, the precision increases by a factor of 10 if $R_{\rm TJO}$ is 3 times bigger or $D_{\rm TJO}$ is 3 times smaller. The exception is $\ln \Theta_{\star}$, for which the precision becomes smaller as $D_{\rm TJO}$ decreases below $\sim 20\ \AU$ for kilometre-sized objects.}

\begin{table*}
\begin{minipage}{140mm}
\caption{Projected $1\sigma$ marginalized uncertainties for parameters}
\label{table:1DUncertainties}
\begin{tabular}{lllcccccc}
\hline
Parameter    & Units & True  & \multicolumn{2}{c}{VERITAS} & \multicolumn{2}{c}{HESS} & \multicolumn{2}{c}{CTA}\\
             &       & Value & Fiducial & ``bright'' & fiducial & ``bright'' & fiducial & ``bright''\\
\hline
\multicolumn{9}{c}{Theorist's parameters (Fisher matrix)}\\
\hline
$r_{\rm min}$      & $\km$      & 0           & 0.097 & 0.0066 & 0.094 & 0.0065 & 0.011 & $7.2 \times 10^{-4}$\\
$v$                & $\kms$     & 30          & 9.5   & 0.65   & 6.6   & 0.45   & 0.24  & 0.016\\
$\theta$           & $\degtext$ & 0           & 12    & 0.83   & 13    & 0.86   & 0.58  & 0.040\\
$t_{\rm min}$      & $\msec$    & 0           & 0.42  & 0.029  & 0.15  & 0.011  & 0.11  & 0.0076\\
$R_{\rm TJO}$      & $\meter$   & 316         & 100   & 7.0    & 70    & 4.8    & 3.1   & 0.23\\
$D_{\rm TJO}$      & $\AU$      & 39.8        & 25    & 1.7    & 18    & 1.2    & 0.62  & 0.042\\
$\Theta_{\star}^a$ & $\muas$    & 14.0 (10.9) & 4.4   & 0.24   & 3.1   & 0.17   & 0.25  & 0.022\\
\hline
\multicolumn{9}{c}{Observer's parameters (Fisher matrix)}\\
\hline
$\ell_F$         & $\km$       & 1.27          & 0.42   & 0.028                & 0.29   & 0.019                & 0.010  & $6.8 \times 10^{-4}$\\
$\nu_{\rm occ}$  & $\sec^{-1}$ & 23.6          & 0.19   & 0.013                & 0.079  & 0.0050               & 0.050  & 0.0034\\
$\rho$           & ...         & 0.248         & 0.0059 & $4.4 \times 10^{-4}$ & 0.0024 & $2.0 \times 10^{-4}$ & 0.0015 & $1.2 \times 10^{-4}$\\
$\rho_{\star}^a$ & ...         & 0.316 (0.247) & 0.022  & 0.0018               & 0.0092 & $7.3 \times 10^{-4}$ & 0.0058 & $4.8 \times 10^{-4}$\\
\hline
\end{tabular}
\\$^a$: The first true value is for the fiducial A0V $V = 12$ star with $\xi = 1$; the value in parentheses is for the bright B5V $V = 10$ star with $\xi = 0$.
\end{minipage}
\end{table*}

{\Blue I list the projected $1\sigma$ marginalized uncertainties on each parameter in Table~\ref{table:1DUncertainties} for the fiducial and bright star occultations.  VERITAS and HESS can measure $r_{\rm min}$, $\theta$, and $t_{\rm min}$ precisely.  The projected $1\sigma$ errors in $v$, $R_{\rm TJO}$, $D_{\rm TJO}$, and $\Theta_{\star}$ are roughly 1/3 the actual values.  Switching to observer's parameters, the reason for the low accuracy is clear: although $\nu_{\rm occ}$, $\rho$, and $\rho_{\star}$ are measured very precisely, $\ell_F$ itself can be measured only to 1 part in 3.  Thus}, even though an event is cleanly detected (with an SNR of 60 in this case), only vague inferences about the TJO's properties can be derived.  

\begin{figure}
\centerline{\includegraphics[width=8.5cm]{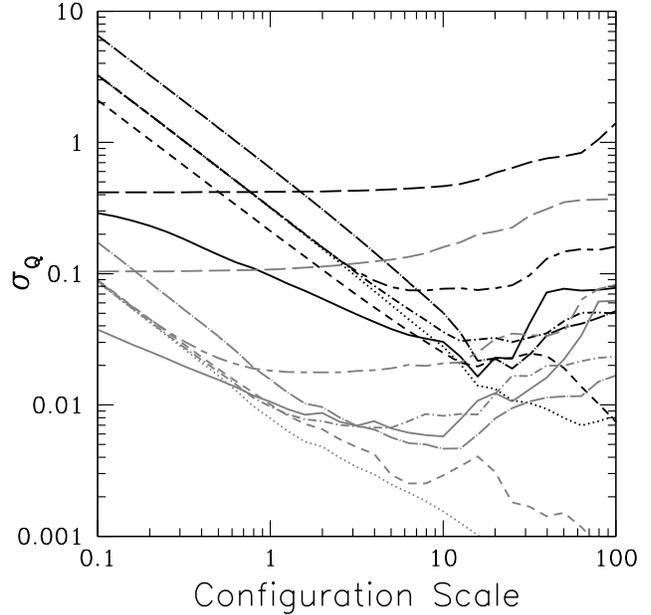}}
\caption{How the errors in estimating each parameter for the fiducial object would vary if the spacings in the array were multiplied by a constant factor.  {\Blue Different line styles correspond to different parameters: $r_{\rm min}$ (solid), $\ln v$ (dotted), $\theta$ (short-dashed), $t_{\rm min}$ (long-dashed), $\ln R_{\rm TJO}$ (short-dashed/dotted), $\ln D_{\rm TJO}$ (long-dashed/dotted), and $\ln \Theta_{\star}$ (short-dashed/long-dashed).} VERITAS {\Blue (black)} is about ten times too small to optimally estimate the object's speed, motion, and distance.  CTA {\Blue (grey)}, however, has a good spacing range for these estimates.  The best configurations are those about the size of the Fresnel radius.\label{fig:ConfigScaleErrors}}
\end{figure}

\begin{figure*}
\centerline{\includegraphics[width=18cm]{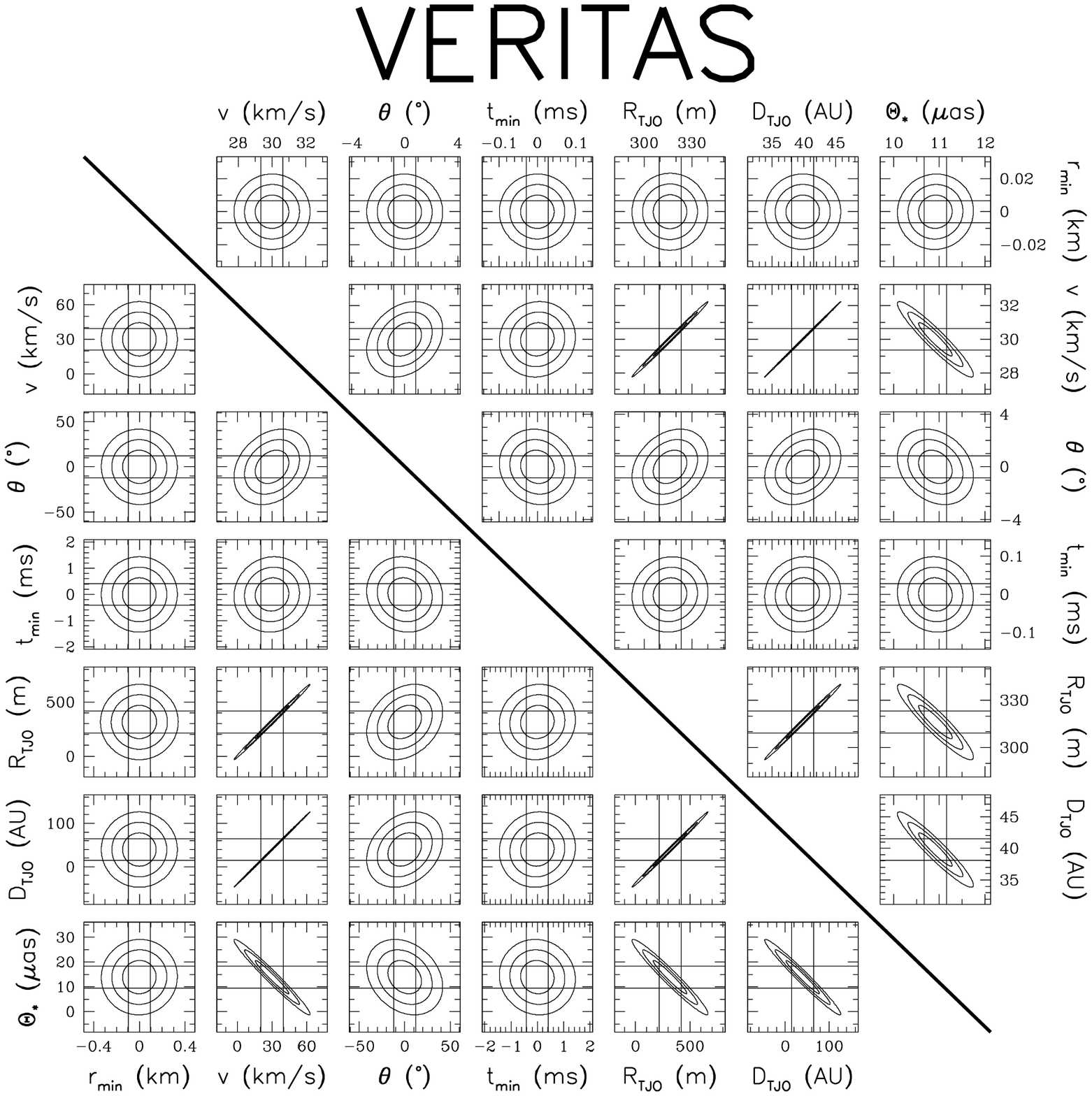}}
\caption{How well VERITAS can estimate occultation parameters for the fiducial occultation.  The vertical and horizontal lines give the $1\sigma$ confidence interval for each parameter.  On bottom left is the fiducial occultation; on top right is the same object occulting the ``bright'' star. \label{fig:GrandEllipsesVERITAS}}
\end{figure*}

In fact, the typical baselines in VERITAS and HESS are $\sim 100\ \meter$, but the Fresnel scale for an object 40 AU away is over 1 km wide.  I show what would happen if the spacings in the VERITAS array were multiplied by a constant scale factor in Figure~\ref{fig:ConfigScaleErrors}.  If VERITAS were ten times larger, it would be roughly the size of the Fresnel scale and could measure most of the occultation parameters precisely for $D_{\rm TJO} = 40\ \AU$.  Despite this limitation, VERITAS and HESS should do well for occultations of bright stars, like the ``bright'' B5V $V = 10$ stars, simply because there are more photons (upper right in Figure~\ref{fig:GrandEllipsesVERITAS}).  In these cases, the errors become small enough to constrain the TJO speed, distance, and size to within 10\%.

The error ellipses for each pair of {\Blue theorist's} parameters for the ``fiducial'' occultation are plotted in Figure~\ref{fig:GrandEllipsesVERITAS} for VERITAS.  The covariances in the parameters are relatively small for this simulated event, except between the speed, radius, distance, and stellar angular radius, where the errors are highly correlated: {\Blue an} occultation {\Blue may be} by a small, close, slow object in front of a small star or a big, far, fast object in front of a large star.  Note that, when the background star is A0V $V = 12$ (bottom left), the Fisher matrix analysis indicates that the 3$\sigma$ confidence ellipses include regions of parameter space where the speed, distance, or radius of the TJO is 0!  Of course, the Fisher matrix analysis is only accurate when the errors in the parameters are much smaller than the parameters themselves -- the very fact that an occultation is detectable means the radius cannot be 0 {\Blue (see the next section)}.

One way to break the degeneracy is if the stellar angular radius is independently known \citep[cf.,][]{Cooray03-Fisher}.  Although VERITAS and HESS are too small to measure the pattern size, they can measure $\rho$ and $\rho_{\star}$ accurately {\Blue (Table~\ref{table:1DUncertainties})}.  By combining prior knowledge of $\Theta_{\star}$ with a derived $\rho_{\star}$, one can calculate the distance of the TJO, and from there, the Fresnel scale, $R_{\rm TJO}$, and $v$.  Using a restricted Fisher matrix where $\Theta_{\star}$ is not counted as a free parameter, I find that VERITAS and HESS are much more powerful at estimating the object's properties if $\Theta_{\star}$ is exactly known.  Specifically, for the fiducial occultation, VERITAS (HESS) estimates $v$ to $2\ \kms$ ($0.7\ \kms$), $R_{\rm TJO}$ to $25\ \meter$ ($9\ \meter$), and $D_{\rm TJO}$ to $5\ \AU$ ($2\ \AU$).

Another way to improve the parameter estimates is if $v$ is smaller, as is the case at quadrature instead of opposition.  Then the theoretical light curve is the same, except that the event is longer.  From the Fisher matrix, we see that the parameter variances scale as $\propto v^{-1}$, or $v^{-3}$ for the speed.  So, for $v = 5\ \kms$, the parameter errors are $\sqrt{6}$ times smaller, except that $v$ is measured $6\sqrt{6}$ times more precisely.  For the fiducial occultation, this alone is enough to estimate $D_{\rm TJO}$ to 10 AU, $R_{\rm TJO}$ to 40 {\Blue metres}, and $\Theta_{\star}$ to $2\ \muas$ with VERITAS.  The price is that occultation events occur more rarely when $v$ is small (Section~\ref{sec:Frequency}).

According to the Fisher matrix analysis, CTA will have impressive capabilities in characterizing TJOs (Figure~\ref{fig:ParamErrorsCTA}).  The sizes and distances of TJOs occulting a fiducial star can be measured to within 10\%, for 1 km radius objects 300 AUs away, or for 100 {\Blue metre} radius objects in the Kuiper Belt.  When CTA observes an occultation of a bright star, it will be able to derive meaningful constraints even for kilometre-radius objects 2000 AU away, or 50 {\Blue metre} radius objects in the Kuiper Belt.  {\Blue As seen in Table~\ref{table:1DUncertainties}, most of the improvement for distance is in measuring $\ell_F$; there are small gains in the precision of $\rho$, $\rho_{\star}$, and $\nu_{\rm occ}$ compared to HESS.}

\begin{figure}
\centerline{\includegraphics[width=8.5cm]{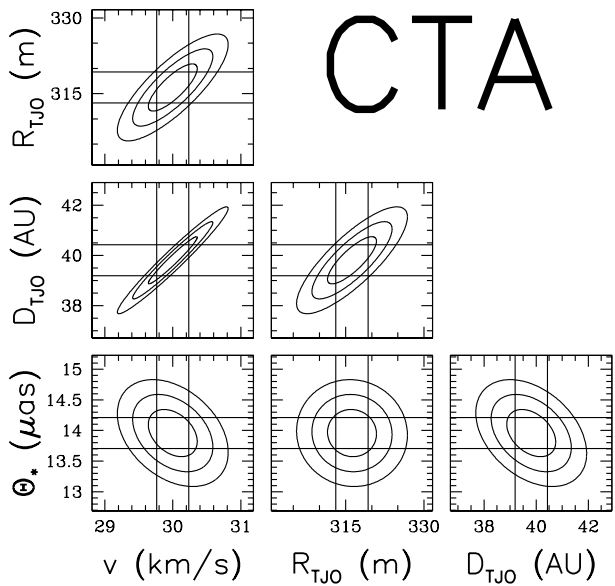}}
\caption{How well CTA can estimate {\Blue certain} occultation parameters {\Blue from} the fiducial occultation.  {\Blue The vertical and horizontal lines give the $1\sigma$ confidence interval for each parameter.}  \label{fig:GrandEllipsesCTA}}
\end{figure}

I show {\Blue a} projected set of error ellipses for the fiducial occultation at CTA in Figure~\ref{fig:GrandEllipsesCTA}.  Even when observing the fainter $V = 12$ fainter star, CTA is capable of constraining the object's parameters to within a few percent.  Although the errors between the TJO speed, radius, and distance are still correlated, {\Blue the covariances are} not as extreme as with VERITAS.  Although many of CTA's baselines are 100 m long, it also includes other baselines spanning up to 3.5 km (Table~\ref{table:IACTArrays}).  Thus, it is larger than the Fresnel scale of a Kuiper Belt object and can measure the size of the diffraction pattern relatively well.  Indeed, I show in Figure~\ref{fig:ConfigScaleErrors} that the E configuration of CTA already approximately has the optimal scale for estimating parameters from Kuiper Belt occultations.  Of course, occultations of bright stars yield even more precise parameter estimates.  For example, according to the Fisher matrix analysis, the radius of the fiducial object can be estimated to within 22 cm.  Naturally, the model itself cannot be this accurate (Section~\ref{sec:ModelComplications}).

Although the planned CTA is well spaced for observing Kuiper Belt Object occultations, it is not well-suited for characterizing objects in the Oort Cloud, simply because the Fresnel scale for these objects is so large.  To improve its parameter estimation capabilities, additional telescopes must be added at a distance of $\ell_F \approx 18\ \km \sqrt{D_{\rm TJO} / 10^4\ \AU}$.  Distant telescopes could prove useful not only for Oort Cloud occultation observations, but for stellar intensity interferometry, another postulated use of the vast photon collection abilities of the array \citep{LeBohec06,Dravins13}.  

\subsection{\Blue Likelihood ratio estimation of parameters for the fiducial model}

{\Blue The projected confidence ellipses include unphysical values according to the Fisher matrix analysis, and a more accurate uncertainty projection is necessary.   A fundamental tool in evaluating models is likelihood.  The likelihood of a model is simply the probability of the data having its observed values under the assumption that the model is true:
\begin{equation}
{\cal L}({\bf \Psi}, {\bf n}) = p({\bf n}, {\bf \Psi})
\end{equation}
for a data vector ${\bf n}$ and a parameter vector ${\bf \Psi}$.  If the data are independent of each other, then the likelihood is the product of the probabilities for each data point.  The parameters can be estimated by finding the model ${\bf \Psi_{\rm ML}}$ with the maximum likelihood.}

{\Blue We can quantify how well a model fits by taking the ratios of likelihoods:
\begin{equation}
\Lambda = 2 \ln \left[\frac{{\cal L} ({\bf \Psi_1}, {\bf n})}{{\cal L} ({\bf \Psi_{\rm ML}}, {\bf n})}\right].
\end{equation}
Confidence bounds can be set by selecting regions in parameter space with $\Lambda$ above some threshold.  According to Wilks' theorem, $\Lambda$ tends to have a $\chi_k^2$ distribution where the number of degrees of freedom $k$ is the number of parameters in the model \citep[e.g.,][]{Cash79}.  The $1\sigma$ confidence region is given by $\Lambda \ge -\chi_k^2 (1)$, the $2\sigma$ contour is given by $\Lambda \ge -\chi_k^2 (2)$ and so on.\footnote{\Blue I checked whether the integrated posterior probability densities within these contours actually had the values of $68\%$ for $1\sigma$, $95.5\%$ for $2\sigma$, and $99.7\%$ for $3\sigma$.  They usually were in rough but not perfect agreement; for example, only $\sim 57$\% of the posterior probability density is within the 1$\sigma$ contour for $\ell_F$ with VERITAS.  These disagreements may simply be due to the coarseness of the grid.}  The Fisher matrix method actually is a simplified case of the likelihood method, which approximates the likelihood as a second-order Taylor series around ${\bf \Psi_{\rm ML}}$.}

{\Blue I estimate the likelihoods at each ${\bf \Psi}$ by assuming that each measurement $n (t_k; l)$ has a value equal to the expected number of photons $n_{\rm sky} + I_{\rm obs}(t_k; l) \bar{n_{\star}}(l)$ and a variance of $\sigma_n^2 (l)$ (equation~\ref{eqn:sigman}).  I also assume that the probability distribution of $n (t_k; l)$ is Gaussian; since the number of photons is large, this should be roughly correct.  Then the log-likelihood is
\begin{align}
\nonumber \ln {\cal L}({\bf \Psi}, {\bf n}) & = -\sum_l \sum_k \left[\frac{(\Delta I_{\rm obs}(t_k; l) \bar{n_{\star}}(l))^2}{2 \sigma_n(l)^2} \right.\\
                                           & + \left. \frac{1}{2} \ln (2 \pi \sigma_n^2(l)) \right].
\end{align}}

\begin{table}
\begin{minipage}{85mm}
\caption{Likelihood parameter grid points}
\label{table:LikelihoodGrid}
\begin{tabular}{lcccc}
\hline
Parameter   & \multicolumn{2}{c}{VERITAS} & \multicolumn{2}{c}{CTA}\\
            & Range & Step & Range & Step\\
\hline
$\displaystyle \frac{\rho}{\rho^{\rm true}}$                            & 0.9 -- 1.1   & 0.01 & 0.975 -- 1.025 & 0.0025\\
$\displaystyle \frac{\rho_{\star}}{\rho_{\star}^{\rm true}}$            & 0.5 -- 1.5   & 0.05 & 0.9 -- 1.1     & 0.01\\
$\displaystyle \frac{\nu_{\rm occ}}{\nu_{\rm occ}^{\rm true}}$          & 0.95 -- 1.05 & 0.01 & 0.99 -- 1.01   & 0.002\\
$\displaystyle \log_{10} \left(\frac{\ell_F}{\ell_F^{\rm true}}\right)$ & -0.7 -- 2.0  & 0.1  & ...            & ...\\
$\displaystyle \frac{\ell_F}{\ell_F^{\rm true}}$                        & ...          & ...  & 0.97 -- 1.03   & 0.006\\
$t_{\rm min}~(\msec)$                      & -2.0 -- 2.0 & 0.4  & -0.4 -- 0.4   & 0.08\\
$\theta~(\degtext)$                        & -50 -- 50   & 10   & -2.5 -- 2.5   & 0.5\\
$r_{\rm min}~(\km)$                        & -0.4 -- 0.4 & 0.08 & -0.04 -- 0.04 & 0.008\\
\hline
\end{tabular}
\end{minipage}
\end{table}

{\Blue The parameter space has seven dimensions, so there is a very large combination of possible values even for one event.  My approach is simply brute force: I calculate likelihood values on a full 7D grid with $\sim 1.8 \times 10^8$ points (see Table~\ref{table:LikelihoodGrid} for details).  I focus on the fiducial event so that the task is manageable.  Brute force methods have been occasionally used before for estimating cosmological parameters \citep[e.g.,][]{Efstathiou99,Tegmark00}.  For larger problems with more events or more parameters, though, a Markov Chain Monte Carlo method or high-order expansions of Fisher matrices \citep*{Sellentin14} may be employed more economically.  In order for brute force to work effectively, the grid must be fine enough to sample ${\bf \Psi}$ with large $\Lambda$ inside each confidence region.  The theorist's parameter set is unsuitable because of the large covariances between variables; the confidence regions are ``tilted'' and slip through a square grid unless a huge number of points are used.  Instead, I use the observer's parameter set.}

{\Blue The resulting confidence bounds are 7D subsets of parameter space, but generally we wish to know the confidence region for just one or two of those parameters.  We must marginalize the likelihoods: to compute the marginal likelihoods of a subset ${\bf \Psi_{\alpha}}$ of the original parameter set, we integrate over the remaining parameters ${\bf \Psi_{\beta}} = {\bf \Psi} / {\bf \Psi_{\alpha}}$:
\begin{equation}
{\cal L}({\bf \psi_{\alpha}}, {\bf n}) = \int_{\bf \Psi_{\alpha} = \psi_{\alpha}} {\cal L}({\bf \Psi}, {\bf n})\ p({\bf \Psi_{\beta}})\ d{\bf \Psi_{\beta}}.
\end{equation}
Likelihood marginalization is implicitly a Bayesian procedure, and requires a prior $p({\bf \Psi_{\beta}})$ over the parameters that are integrated out \citep[e.g.,][]{Freeman99}.  I assume flat uniform priors for all of the natural parameters \citep[cf.,][]{Efstathiou99}.}

\begin{figure*}
\centerline{\includegraphics[width=8.5cm]{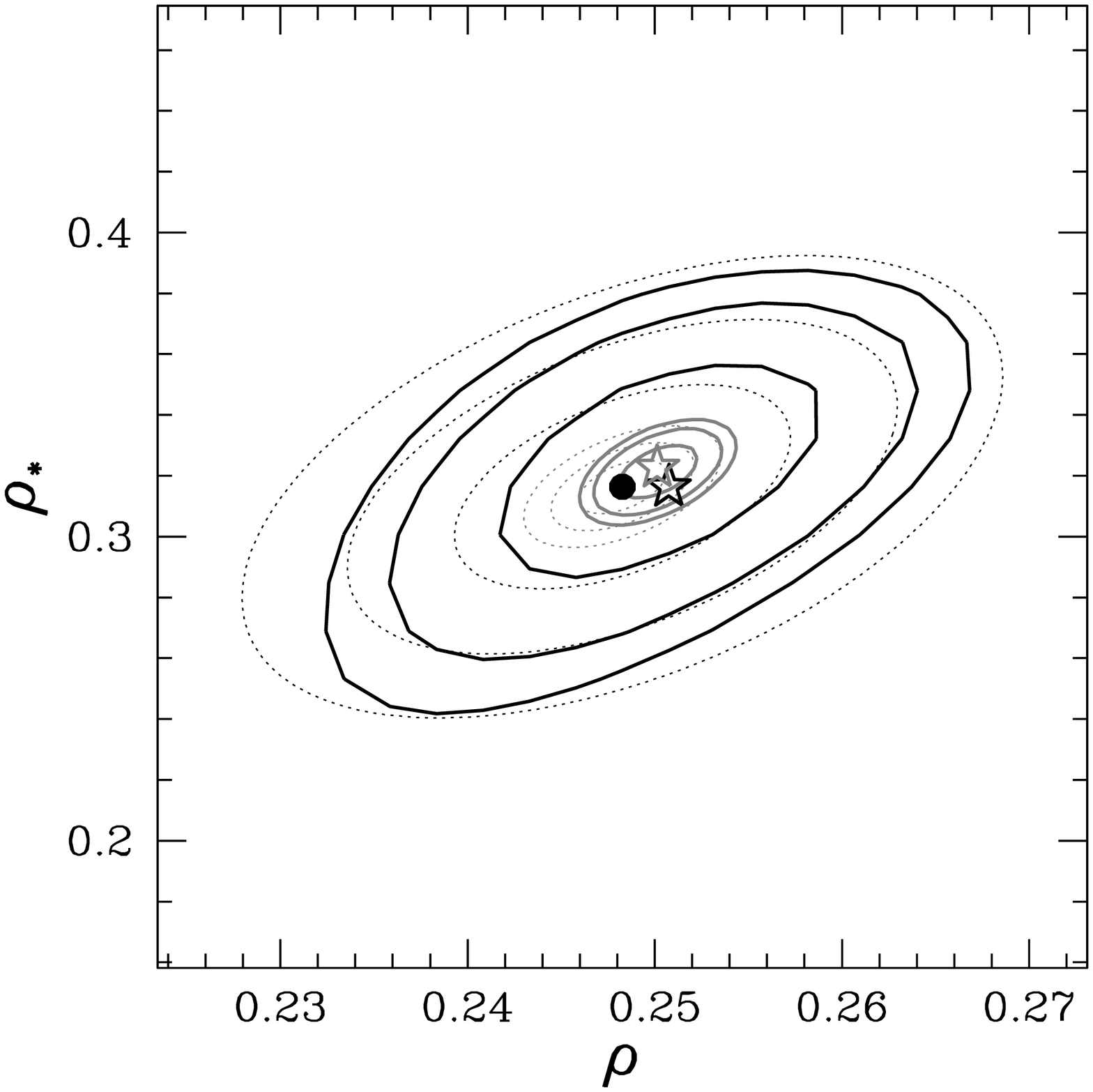}\ \includegraphics[width=8.5cm]{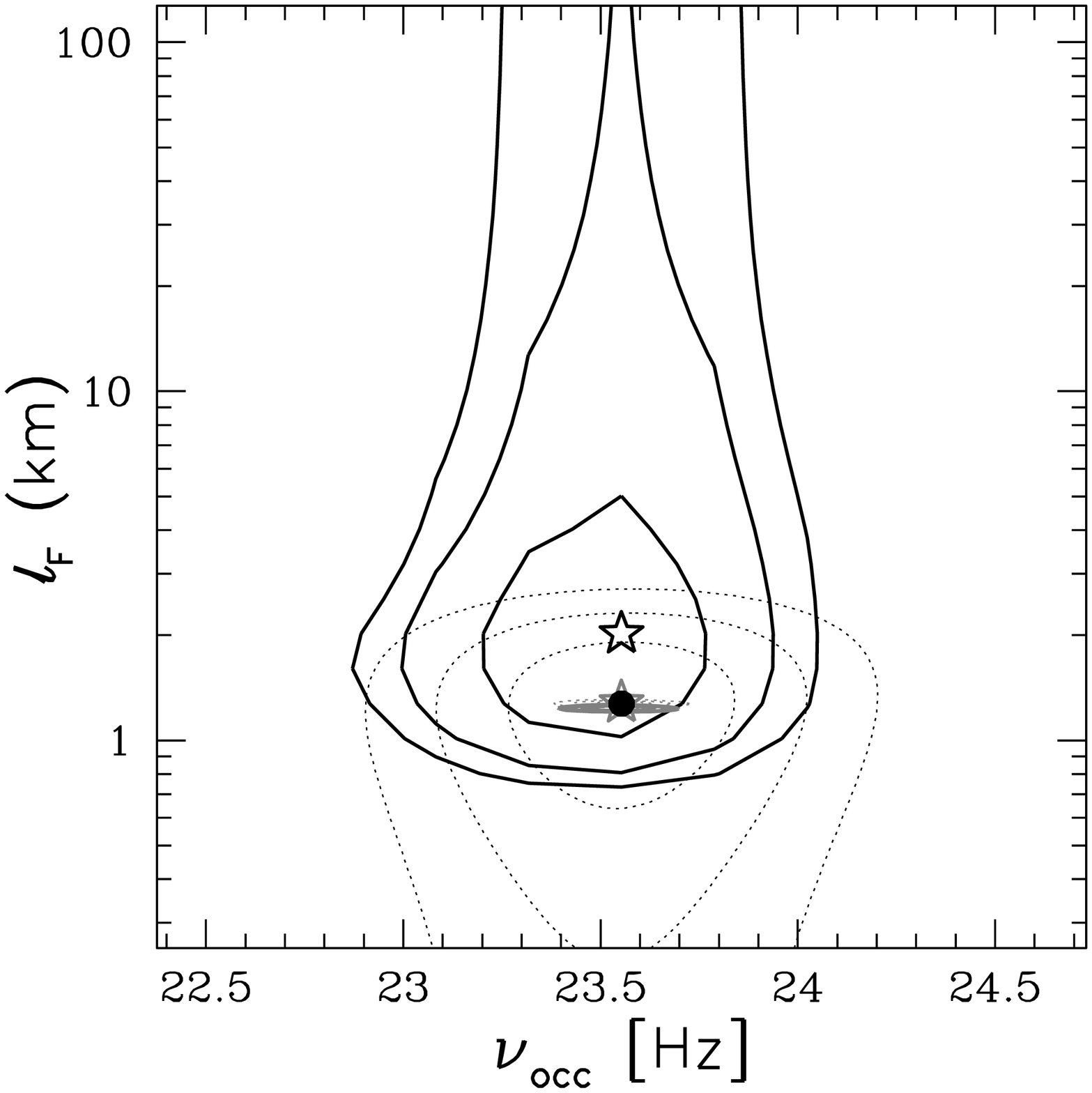}}
\caption{\Blue Projected uncertainties ($1\sigma$, $2\sigma$, $3\sigma$) using the likelihood ratio method (solid lines), for $\rho$ and $\rho_{\star}$ (left), and $\nu_{\rm occ}$ and $\ell_F$ (right).  The projections for VERITAS are in grey, while those for CTA are in black.  For comparison, I show the Fisher matrix projected uncertainties as the thin dotted ellipses.  The big black dot is the actual model parameters, while the stars are the maximum likelihood parameter estimates for each telescope array. \label{fig:FidLikelihoodBounds}}
\end{figure*}

{\Blue The coarseness of the grid limits the precision of the uncertainty estimates, but I find that the projected 1$\sigma$ marginalized uncertainties are comparable to those found with the Fisher matrix formalism.  The exceptions are for $r_{\rm min}$, for which the uncertainties are $1.6$ times bigger than the Fisher matrix uncertainties, and $\ell_F$, which has very weak upper bounds only.  Some projected uncertainty ellipses for the fiducial model are shown in Figure~\ref{fig:FidLikelihoodBounds}.  I find that the sizes of the confidence ellipses found with the Fisher matrix method (dotted) are similar to the regions found with the likelihood ratio method (solid), except when estimating $\ell_F$.  Although VERITAS can set a lower bound on the Fresnel scale, it cannot effectively distinguish between models with Fresnel scales that are much bigger than the array.  As expected from the discussion in Section~\ref{sec:FisherResults}, CTA can constrain all parameters effectively for the fiducial occultation.}

{\Blue One minor difference with the fiducial model is that the maximum likelihood parameter set (stars in Figure~\ref{fig:FidLikelihoodBounds}) is not the same as the true parameter set (big dots).  As a result, the confidence regions are shifted away from the true parameter set to some extent.}

\subsection{What else might be measurable?}
\label{sec:ModelComplications}
The Fisher matrix analysis results in some extremely precise parameter estimates, especially with the CTA.  But although the Fisher matrix analysis implies high precision for a given model, that does not mean the model is accurate.  In fact, the models I use are highly idealized.

One assumption that may be relaxed is that of a perfectly spherical TJO.  \citet{Roques87} discussed the diffraction patterns of nonspherical bodies.  For an object larger than the Fresnel scale, the pattern is simply a geometrical shadow, and the TJO's shape can be inferred by the time that the star is obscured at each telescope (as done for larger asteroids and TJOs).  For objects smaller than the Fresnel scale, the shape of the object does not affect the size of the diffraction pattern.  But \citet{Roques87} shows that the shape of the object affects the ``ringing'' of the diffraction pattern.  There has not been a thorough study of how the ringing depends on shape, but it potentially encodes information that IACTs can exploit to infer the object shape. \citet{Young12} demonstrated how the diffraction pattern of the \citet{Schlichting09} TJO varies with oblateness.  The different shapes in fact break the pattern's radial symmetry, as well as affecting the details of the ringing.  As IACT arrays can probe the two dimensional structure of diffraction patterns, they might constrain TJO shape.  Detailed studies of their performance can be studied with the methods of \citet{Roques87} or \citet{Young12}.

The other obvious idealization in the model is the assumption of a perfectly spherical and uniform background star.  Real stars deviate from these assumption in three major ways.  First, they can be oblate due to their rapid rotation.  Secondly, they are limb-darkened, with the centre of the stellar disc appearing hotter and brighter than the edges.  Finally, they can have large starspots.  

\section{Frequency of Occultations at Cherenkov Telescopes}
\label{sec:Frequency}
A calculation of the event rate requires the sensitivity of the IACT array to a given event, the number of target stars, and a rate distribution for each event.

There are many distinct populations of TJOs.  These span a vast range of sizes, ranging from dust grains to Pluto and Eris.  The size distribution of a population of TJOs is often described by a broken power law.  At larger radii, the number of objects increases extremely steeply as the radius decreases.  But at some scale, the size distribution function turns over, and the number of objects increases more slowly as radius continues to decrease, with $dN/dR \propto R^{-q}$.  The turnover to a $q = 3.5$ spectrum has been seen in populations including Kuiper Belt objects and irregular satellites (\citealt*{Sheppard06-Moons}; \citealt{Schlichting12}).

The volume filled by a population can also cover a large range in distances: the Centaurs range from 10 to 30 AU away, and the Oort Cloud may reach from 3000 to $10^5$ AU away.  I assume that the distance distribution function is a power law $dN/dD_{\rm TJO} = D_{\rm TJO}^{-\gamma}$ truncated at a minimum and maximum distance $D_{\rm min}$ and $D_{\rm max}$.

Nor are the TJOs necessarily evenly distributed across the sky.  The simplest possible assumption is that they uniformly cover some solid angle $\Omega$.  Observations generally constrain the integrated number of objects with radii above some value $R_0$ that is smaller than the power law break.  The number of these objects per unit solid angle is ${\cal S}_0$.  Combining the distance and size dependences, the normalized distribution function of a TJO population is
\begin{equation}
\frac{dN}{d\ln D_{\rm TJO} d\ln R_{\rm TJO}} = \frac{(1 - \gamma)(q - 1) (R / R_0)^{1-q}}{(D_{\rm max}/D)^{1-\gamma} - (D_{\rm min}/D)^{1-\gamma}} {\cal S}_0.
\end{equation}

\begin{figure}
\centerline{\includegraphics[width=8.5cm]{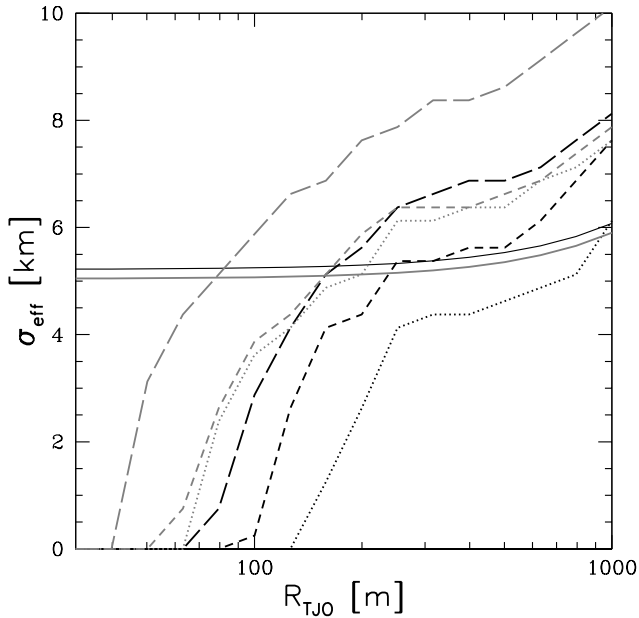}}
\caption{The effective cross sections of diffraction patterns of an object 40 AU away as observed with VERITAS (dotted), HESS (short-dashed), CTA (long-dashed).  The black curves are for the fiducial star and the grey curves are for the ``bright'' star.  The solid lines are the cross sections as according to {\Blue N07}.\label{fig:CrossSections}}
\end{figure}

We can think of the diffraction patterns as moving around on the celestial sphere, each covering a small part of it.  Then the rate of detectable occultations of a given star is
\begin{equation}
\label{eqn:OccultationRate}
\Gamma_{\star} = \int_{D_{\rm min}}^{D_{\rm max}} \int_0^{\infty} \frac{\sigma_{\rm eff} v}{D_{\rm TJO}^2} \frac{dN}{dR_{\rm TJO} dD_{\rm TJO}} dR_{\rm TJO} dD_{\rm TJO}.
\end{equation}
I use $v = 30\ \kms$, appropriate for TJOs at opposition.  The effective cross section $\sigma_{\rm eff}$ of the diffraction pattern is the width of the parts of the pattern that are detectable:
\begin{equation}
\sigma_{\rm eff} = 2 \int_0^{\infty} \left[\frac{\nu}{\sigma_{\nu}} \ge \sqrt{2}\ {\rm erfc}^{-1} (2p)\right] dr_{\rm min}.
\end{equation}
Note that it is not in general equal to, for example, the Fresnel scale or the cross section given in {\Blue N07}, because IACT arrays are sensitive to the outer fringes of the diffraction pattern.  I plot the effective cross sections to an occultation by the fiducial object as observed by the IACT arrays in Figure~\ref{fig:CrossSections}.  The $\sigma_{\rm eff}$ reach values that are about twice as high as the {\Blue N07} width.  Since $\sigma_{\rm eff}$ depends on the detectability of the outer fringes, it depends on the magnitude and type of the observed star, and which IACT array is observing.  

IACT arrays have wide fields of view, typically containing hundreds of stars, so the event rate observed by the IACT array is
\begin{equation}
\Gamma_{\rm IACT} = \frac{{\cal S}_{\star} \Gamma_{\star}}{(\pi/4) \Theta_{\rm FOV}^2},
\end{equation}
where ${\cal S}_{\star}$ is the number of stars per unit solid angle and $\Theta_{\rm FOV}$ is the IACT field of view. Formally, I need to integrate over stellar magnitude, type, and extinction.  I consider a simpler approximation with two populations of stars: those with $V \le 12$, and those that are comparable to the ``bright'' star.  For the former case, I assume that the sensitivity of IACTs to the fiducial star is typical of a $V \le 12$ star.  I justify this by noting that the sensitivity of IACT arrays is comparable for dwarfs and giants earlier than K type (Figure~\ref{fig:DetectabilityStarV}). In addition, I compare to the stellar angular size distribution derived by \citet{Cooray03-Rates}.  They find that roughly half of $V = 12$ stars have an angular radius that is less than $28\ \muas$, or twice the angular radius of the fiducial star.  

The second case of the ``bright'' stars is more complicated, since B5V $\xi = 0$ stars are at the small-angular size tail of the angular size distribution.  According to the \citet{Cooray03-Rates} model, essentially no stars have angular sizes that small.  Only about 1 in 15 are less than twice as big as viewed from Earth.  On the other hand, the optimal sensitivity to these stars is not just because of their small size, but also their low extinctions, which allows through the blue light that PMTs are most sensitive to.  I simply divide the total number of stars brighter than $V = 10$ by 15 to arrive at the number of ``bright'' stars, but a more rigorous model is desirable.  

The integrated number of stars brighter than each magnitude as a function of Galactic latitude is given in \citet{Bahcall80}.  I use the worst case of the Galactic Poles, but note that the density at $|b| \approx 20^{\circ}$ is $\sim 3$ times larger \citep{Bahcall80}.  There are ten $V=12$ stars per square degree, and $2/15 = 0.13$ ``bright'' stars per square degree.   I adopt a field of view of $5^{\circ}$ and assume that the sensitivity is uniform across the field, although angular resolution actually degrades towards the edges.

I list the expected occultation rates for several populations of TJOs in Table~\ref{table:EventRatesPerArray}.  

\begin{table*}
\begin{minipage}{140mm}
\caption{Estimated total event rates total for sub-kilometre TJO populations}
\label{table:EventRatesPerArray}
\begin{tabular}{llcccccc}
\hline
Population   & Distribution  & \multicolumn{2}{c}{VERITAS} & \multicolumn{2}{c}{HESS} & \multicolumn{2}{c}{CTA}\\
             & & \multicolumn{2}{c}{${\rm max}[\Gamma_{\rm IACT}^{-1} ({\rm hr})]$} & \multicolumn{2}{c}{${\rm max}[\Gamma_{\rm IACT}^{-1} ({\rm hr})]$} & \multicolumn{2}{c}{${\rm max}[\Gamma_{\rm IACT}^{-1} ({\rm hr})]$} \\
             & & $V \la 12$ & ``bright'' & $V \la 12$ & ``bright'' & $V \la 12$ & ``bright''\\
\hline
KBOs                                   & S12, $q = 3.5$ & 10    & 100     & 4     & 80     & 2      & 20\\
Centaurs                               & S00, $q = 3.5$ & 700   & 7000    & 200   & 5000   & 100    & 1000\\
                                       & S00, $q = 4.0$ & 60    & 400     & 10    & 200    & 7      & 50\\
Scattered-Disc Objects                 & T00, $q = 3.5$ & 300   & 2000    & 90    & 2000   & 50     & 500\\
Oort Cloud Core                        & C13, $q = 3.5$ & 20000 & 100000  & 6000  & 90000  & 3000   & 30000\\
                                       & C13, $q = 4.0$ & 1000  & 5000    & 300   & 3000   & 100    & 800\\
Hills (Oort) Cloud                     & O10, $q = 3.5$ & ...   & $10^6$  & ...   & 400000 & 500000 & 70000\\
                                       & O10, $q = 4.0$ & ...   & 800000  & ...   & 300000 & 400000 & 50000\\
Uranus Trojans                         & A13, $q = 3.5$ & 60000 & 500000  & 20000 & 400000 & 9000   & 100000\\
                                       & A13, $q = 4.0$ & 5000  & 30000   & 1000  & 20000  & 600    & 5000\\
Neptune Trojans                        & S10, $q = 3.5$ & 300   & 2000    & 90    & 2000   & 50     & 500\\
                                       & S10, $q = 4.0$ & 20    & 90      & 4     & 60     & 2      & 10\\
Uranus Outer Satellites                & S06            & 900   & 9000    & 300   & 6000   & 100    & 2000\\
                                       & $q = 3.5$      & 600   & 5000    & 200   & 4000   & 90     & 1000\\
                                       & $q = 4.0$      & 80    & 500     & 20    & 300    & 9      & 70\\
Neptune Irregular Satellites           & S06            & 2000  & 10000   & 500   & 10000  & 300    & 3000\\
                                       & $q = 3.5$      & 200   & 2000    & 70    & 1000   & 40     & 400\\
                                       & $q = 4.0$      & 20    & 100     & 5     & 70     & 2      & 20\\
\hline
\end{tabular}
\\I assume a field of view of $5^{\circ}$ for all telescopes.  I also assume that occultations of all $V = 12$ stars are as detectable as for a $V = 12$, $\xi = 1$ A0V star, and that 7\% of all occultations of $V = 10$ stars are as detectable as for a $V = 10$, $\xi = 0$ B5V star.\\
S12: \citet{Schlichting12}; S00: \citet{Sheppard00}; T00: \citet*{Trujillo00}; C13: \citet{Chen13}; O10: \citet{Ofek10}; S06: \citet{Sheppard06-Moons}; S10: \citet{Sheppard10-NTrojanSizes}; A13: \citet{Alexandersen13}\\
\end{minipage}
\end{table*}

\subsection{Kuiper Belt}
The greatest reservoir of minor bodies within 100 AU of the Sun is the Kuiper Belt.  The ``classical belt'' of these objects includes those with nearly circular orbits in the ecliptic with semimajor axes of $\sim 45\ \AU$.  Kuiper Belt Objects (KBOs) can also be in orbital resonance with Neptune, as Pluto is.  KBOs are the only population of TJOs for which there have been convincing occultations observed \citep{Schlichting09,Schlichting12}.

The size distribution function of KBOs is fairly well known.  At the large object end, it is inferred from direct observations of their brightnesses \citep*[e.g.,][]{Trujillo01}.  The many surveys for KBO occultations have placed strong limits on the abundance of objects with radii between $\sim 10\ \meter$ and 1 km \citep{Bickerton08,Bianco09,Bianco10,Zhang13}.  The two occultations detected by HST have finally pinned down the abundance of small KBOs \citep{Schlichting09,Schlichting12}.  The evidence points to a broken power law distribution, with an exponent $3.5 \la q \la 4.0$ for radii up to 45 km, above which the distribution steepens.

Since large KBOs have low inclinations, it is thought that small KBOs are also concentrated towards the ecliptic.  According to \citet{Schlichting12}, the number density of KBOs with $R_{\rm TJO} \ga 250\ \meter$ is ${\cal S}_0 \approx 4.4 \times 10^6\ \degtext^{-2}$ if the KBOs are uniformly distributed at ecliptic latitudes $|i| \le 20^{\circ}$.  I use this surface density and $q = 3.5$.  The KBOs are assumed to be uniformly distributed between 42 and 45 AU with $\gamma = 0.0$.

I find that IACT arrays detect KBO occultations frequently, of order once per $\la 10$ hours for $V \le 12$ stars when observing fields within $|i| \le 20^{\circ}$ at opposition, using the star counts near the Galactic Poles (Table~\ref{table:EventRatesPerArray}).  The large event rate is due to their sensitivity to the numerous small objects, combined with the large field of view.  High quality occultations of the ``bright'' stars occur about once every 100 hours per observations of fields near the ecliptic.  Note that the region inhabited by KBOs, $|i| \le 20^{\circ}$, covers about 1/3 of the celestial sphere.

\subsection{Centaurs}
The Centaurs are former KBOs that have been scattered into unstable orbits between Jupiter and Neptune. Since the lifetime of the Centaur orbits is only a few Myr, there are a lot fewer of these objects than KBOs at any given time.  However, they are nearer to Earth and more easily detected.  

\citet{Sheppard00} extrapolated the observed number of Centaurs down to a radius of $R_0 = 1\ \km$.  For a uniform inclination distribution with $|i| \le 30^{\circ}$, they estimate the surface density is ${\cal S}_0 = 150\ \degtext^{-2}$ with $q = 3.5$ or $600\ \degtext^{-2}$ with $q = 4.0$.  The distance distribution function has a slope $\gamma = 1.3$; where I take $D_{\rm min} = 10\ \AU$ and $D_{\rm max} = 30\ \AU$ \citep{Sheppard00}.

If the Centaurs have a shallow size distribution of $q = 3.5$, then observable occultations are fairly rare, occurring once every few hundred hours at opposition for $V = 12$ stars in fields near the ecliptic.  HESS and VERITAS typically accumulate 1000 hours of dark sky observations per year, so we may expect these occultations to be observed {\Blue at most} about once a year.

\subsection{Scattered-Disc Objects}
Just as some KBOs have been scattered inwards to become Centaurs, others have been scattered outwards.  These scattered disc objects (SDOs) have perihelia near 40 AU but big eccentricities and inclinations, with some ranging out to 100 AU.  Whether they were ejected from the Kuiper Belt early in the Solar System's history or through the present time is unclear.  In this paper, I simply consider SDOs to be a population of objects that spans $D_{\rm min} = 50\ \AU$ to $D_{\rm max} = 100\
 \AU$ with $\gamma = 0$.

\citet{Trujillo00} estimates that there are 30000 SDOs with perihelia less than 36 AUs and radii greater than 50 km.  According to \citep{Gomes08}, the total number of large SDOs with any perihelion is twice this number.  If the SDOs are distributed isotropically on the sky, this gives ${\cal S}_0 = 4800\ \sr^{-1}$ at $R_0 = 50\ \km$.  I extrapolate to smaller sizes by assuming $q = 3.5$.

SDO occultations are rarer than KBO occultations, but appear to be more common that Centaur occultations if $q = 3.5$ for both populations.  IACT arrays observe SDO occultations of $V = 12$ stars roughly once every few hundred hours (Table~\ref{table:EventRatesPerArray}).

\subsection{Oort cloud}
The greatest reservoir of minor icy bodies, the Oort Cloud is thought to lie well beyond direct observation, at 1000 to 100000 AU away.  The 10000 AU aphelia of long period comets is an indirect hint that objects exist at these distances.  The Oort Cloud was likely formed by the outer planets as they cleared out the early Solar System.  For a long time, the Oort Cloud was thought to have two populations: an outer ``classical'' Oort cloud ($a \ga 20000\ \AU$) susceptible to perturbations of other stars, and an inner Hills cloud within 20000 AU that is generally unperturbed except by the closest stellar passages (\citealt{Hills81}; \citealt*{Duncan87}).  The discovery of 90377 Sedna, a large object with a perihelion of 76 AU and an aphelion of 480 AU, complicated this picture \citep{Brown04}.  The existence of Sedna and similar bodies may indicate that interactions with other stars in the Sun's birth cluster shaped the Oort Cloud, which may be actually compressed to within a few thousand AU of the Sun \citep{Fernandez00,Morbidelli04}. 

I consider two populations of Oort Cloud objects.  The first is a ``core'' of Oort Cloud bodies with orbits similar to Sedna.  Previous limits on this population were set by TAOS \citep{Wang09}.  According to \citet{Chen13}, there are 11000 Oort Cloud Core objects with radii greater than $R_0 = 20\ \km$.  If they are isotropically distributed across the sky, this gives ${\cal S}_0 = 880\ \sr^{-1}$.  I assume the Oort Cloud Core spans $D_{\rm min} = 200\ \AU$ to $D_{\rm max} = 1000\ \AU$ with $\gamma = 0$, and I consider $q = 3.5$ and $q = 4.0$.  

The second is the Hills Cloud, as considered by \citet{Ofek10}.  There are of order $10^{12}$ classical Oort Cloud objects with radii greater than $R_0 = 1\ \km$, although the true number is very uncertain.  Given an isotropic distribution on the sky, ${\cal S}_0 = 8 \times 10^{10}\ \sr^{-1}$.  \citet{Ofek10} also quotes a distance distribution of $\gamma = 3.5$ between $D_{\rm min} = 3500\ \AU$ and $D_{\rm max} = 50000\ \AU$ from \citet{Duncan87}.  

Unfortunately, although IACTs are capable of detecting these objects, their distance works against them (equation~\ref{eqn:OccultationRate}).  Observable occultations of the Oort Cloud Core bodies require tens of thousands of hours of observations if $q = 3.5$, which would take several decades of observational time.  If $q = 4.0$, then only a few hundred hours with HESS or CTA may suffice.  Occultations by classical Oort cloud objects are even rarer, with one in an IACT array's field of view every few hundred thousand hours, requiring centuries of observation.

\subsection{Outer Planet Trojans}
A minor body can co-orbit with a planet if it is trapped at the L4 or L5 Lagrangian points, located $60^{\circ}$ ahead or behind respectively of the planet.  {\Blue Their orbits} can remain stable for longer than the age of the Solar System in some cases.  The largest Trojan population in the Solar System is associated with Neptune, which actually outnumbers the main asteroid belt \citep{Sheppard06-NTrojans}.  Neptune Trojans are observed at both the L4 and L5 points \citep{Sheppard10-NTrojanL5}.  Although not nearly as numerous as Neptune Trojans, Trojan bodies are also known to co-orbit with Earth, Mars, Jupiter, and Uranus (e.g., \citealt*{Jewitt00}; \citealt*{Connors11}; \citealt{Alexandersen13}); Saturn may have a small Trojan population as well \citep[e.g.,][]{Nesvorny02}. 

The number of large Neptune Trojans is known from surveys presented in \citet{Sheppard06-NTrojans}, \citet{Sheppard10-NTrojanSizes}, and \citet{Sheppard10-NTrojanL5}.  The current estimate is that there are 400 bodies with radii greater than $R_0 = 50\ \km$ \citet{Sheppard10-NTrojanSizes}, with roughly half at L4 and half at L5 \citet{Sheppard10-NTrojanL5}.  The size distribution appears to break from $q = 5$ to smaller $q$ below radii of 45 km, just as for KBOs \citep{Sheppard10-NTrojanSizes}.  The inclinations of Neptune Trojans can be up to $|i| \la 25^{\circ}$, and each population spans $30^{\circ}$ in ecliptic longitude.  With two populations, this means that the surface density of large Trojans is ${\cal S} = 0.13\ \deg^{-2}$.  I then consider extrapolations down to sub-kilometre sizes with $q = 3.5$ and $q = 4$.

I find that, when IACT arrays observe fields where the Trojans are at opposition, a detectable occultation of a $V = 12$ star occurs once every $\la 100$ hours if $q = 3.5$ and once every $\la 10$ hours if $q = 4.0$.  These rates are only $\sim 10$ times slower than of the far more numerous KBOs because the Trojans are confined to a relatively small region of the sky, enhancing their surface density.

The trailing (L5) Neptune Trojans present a very interesting opportunity during the next few years.  These bodies are currently in Sagittarius, a region that HESS frequently observes because it also hosts the Galactic Centre.  In addition, this region has a much higher number of target stars than the Galactic Pole.  Although the large density of stars makes Trojans difficult to detect with conventional observations, it is preferable for occultation detections.  Therefore, the likelihood of serendipitous discoveries by HESS is relatively high. 

The orbits of Neptune Trojans are generally stable, but most possible Saturn and Uranus Trojan orbits are unstable.  As a result, the populations of Saturn and Uranus Trojans are much smaller, and no Saturn Trojans have been detected \citep{Nesvorny02}.  A few dynamically stable regions do exist and may host Trojans \citep*{Dvorak10}.  In addition, Uranus can temporarily capture Centaurs into Trojan orbits \citep{Alexandersen13}. 

The recent discovery of a single Uranus Trojan confirms that these objects exist \citep{Alexandersen13}.  According to simulations also done by \citet{Alexandersen13}, roughly 0.4\% of Centaurs at any given time are trapped into Uranus' Trojan regions.  In computing the Uranus Trojan occultation rate, I therefore use the Centaur size distribution function, but multiplied by 0.004.  I also assume that Uranus' Trojans are trapped in two regions of $1500\ \degtext^{-2}$.  

As is clear from Table~\ref{table:EventRatesPerArray}, the prospects for detecting occultations by Uranus Trojans are dim.  Detectable occultations of a $V = 12$ star at opposition occur only once every $\sim 10^4$ hours of observations of Uranus' Trojan regions if $q = 3.5$.  Even if $q = 4.0$, $\sim 1000$ hours of observations are necessary.  Unless IACT arrays dedicate a year of their observation time or more specifically Uranus' Trojan regions, they are unlikely to set interesting constraints on the population.

\subsection{Irregular satellites of Uranus and Neptune}
Ground-based observations of the past 15 years have revealed swarms of satellites orbiting tens of millions of km from each of the gas and ice giants.  These are the irregular satellites, and they were probably originally captured from the Kuiper Belt.  The collisional time scale for the irregular satellite swarms is relatively short, resulting in a shallow size distribution \citep{Bottke10}.  There are relatively few irregular satellites around each planet, but unlike all of the other populations, they are all concentrated into a tiny sky area.  Stable orbits for irregular satellites can be found within 0.7 Hill radii of the planet \citep{Sheppard06-Moons}.  In other words, the irregular satellites are within about $1^{\circ}$ of the planet. This boosts the odds of seeing an occultation by a sub-kilometre satellite \emph{if} the IACT array specifically observes the region near the planet itself.  There may be additional populations of quasi-satellites orbiting at several times the Hill radius, especially around Uranus and Neptune, although no such objects are known {\Blue yet} \citep{Shen08}.

\citet{Sheppard06-Moons} compares the size distribution functions of the gas giants.  In contrast to the usual pattern of size distribution, Jupiter, Saturn, and Uranus have shallow $dN/dR$ at large radii.  Surveys of the satellite systems of Jupiter and Saturn are complete enough to discern a break in the size distribution at 5 km.  Below this point, the size distribution steepens and is consistent with $q = 3.5$.  The simulations of \citet{Bottke10}, in which these irregular satellite systems are the ruins of much larger systems after collisional grinding, imply that $q < 3.5$ below 5 km.  Neptune's irregular satellites, at least those outside the orbit of Nereid, appear to have a steeper size distribution ($q \sim 4$) to the limits of completeness, 20 km, but the statistics are very low \citep{Sheppard06-Moons}.  

There are 9 known irregular satellites of Uranus, with radii above 7 km.  I consider three possible size distributions.  The first (the S06 distribution) is that the size distribution is similar to those of the Jovian and Saturnian system; with $q = 2$ above $R_0 = 5\ \km$ and $q = 3.5$ below that point.  The second is a $q = 3.5$ distribution at all radii below $R_0 = 7\ \km$, and the third is $q = 4.0$ below $R_0 = 7\ \km$.  I find the occultation rate depends strongly on which distribution is more accurate.  If the S06 or $q = 3.5$ distribution holds, then several hundred hours of observation of the region around Uranus are necessary to detect an occultation of a $V = 12$ star at opposition.  However, if $q = 4.0$, the rates are more optimistic: a few tens of hours are sufficient to detect an occultation.  

The Neptunian satellite system was catastrophically altered by the capture of Triton \citep[e.g.,][]{Cuk05}, and it is unclear whether Triton and Nereid should count as irregular satellites.  There are 5 definitely irregular satellites with $R_0 = 20\ \km$, and I also include Nereid for a total of 6.  I again consider the S06 distribution with a break to $q = 3.5$ at $R_0 = 5\ \km$, and $q = 3.5$ and $q = 4.0$ distributions with $R_0 = 20\ \km$.  Detectable occultations are even rarer for Neptunian irregular satellites than for Uranian irregular satellites if the S06 distribution always holds, with about a thousand hours of observation time dedicated to Neptune required.  If $q = 3.5$ for all radii, though, then only about 100 hours of observation time of Neptune is necessary to detect an occultation of a $V = 12$ star.  For $q = 4.0$, which is currently consistent with data, then $\sim 10$ hours of observations of Neptune yield an occultation by an irregular satellite.

There is no TeV gamma-ray science case for dedicated observations of Uranus and Neptune with IACTs.  None the less, a few nights of dark sky observations can rule out $q = 4.0$ for small satellites.  A limit on $q = 4$ is especially interesting for Neptune, where the satellite size distribution is very poorly understood \citep{Sheppard06-Moons}. 

\section{Conclusions}
IACT arrays are in many ways superior instruments for studying TJO occultations.  Their telescopes are enormous with great light collection areas and relatively little scintillation noise; the telescopes come in arrays, allowing a veto of scintillation noise and potentially breaking the velocity-size-distance degeneracy; the photon sampling rates are essentially infinite from the occultation perspective; and their fields of view ($\sim 5^{\circ}$) are large enough to observe dozens of stars at once.  The disadvantages are large amounts of sky noise because of the poor angular resolution, too short baselines for TJO parameter estimation, and relatively low duty cycle because of moonlight.  {\Blue In addition, the current electronics of the PMTs in IACTs may not be suited for continuous photometry; if they do not use the standard readout modes, there can be deadtime within the integration time (Benbow, private communication).}

I have shown that the already extant VERITAS and HESS arrays are in principle capable of detecting occultations by sub-kilometre TJOs.  The two KBO occultations detected by the HST would have been easily detected by either of these arrays.  HESS, with its central 28 {\Blue metre} telescope, can detect more objects than VERITAS, but has comparable performance for parameter estimation.  {\Blue The greatest obstacle for these arrays is estimating $\ell_F$.}  I estimate that they observe a KBO occultation of a $V = 12$ star once every $\la 10$ hours of near-ecliptic observations. The arrays {\Blue could} also detect a number Centaurs and Scattered Disc Objects during the course of a year, and are capable of setting interesting constraints on the number of Neptune Trojans and irregular satellites.  Much rarer occultations of blue $V = 10$ stars occur at $\la 10\%$ of the rate of the $V = 12$ stars, but observations of these events are worthwhile, as they allow for precise determinations of the TJO size and distance.  

The future CTA will be an excellent facility for characterizing KBO occultations as it will include kilometre-long baselines.  It should be able to detect objects with radii smaller than 100 {\Blue metres} in the Kuiper Belt, allowing it to explore the same parameter space as X-ray occultation searches \citep[e.g.,][]{Liu08,Chang11}, but with many more targets.  CTA also can detect 1 km radius objects out to several thousand AUs, into the inner reaches of the Oort Cloud.  Even for an occultation of a $V = 12$ star by a 250 {\Blue metre} radius KBO, CTA will be able to constrain properties to within a few percent.

Given the apparent great potential of these instruments, it would be worth considering more complex models.  First, a better model of scintillation noise is desirable, as its properties for large telescopes are poorly understood, and it may be correlated on relevant time-scales \citep{Bickerton09}.  Second, the IACTs are so powerful that the assumptions of a uniformly bright star and a spherical object may be insufficient.  A study of whether these instruments can constrain limb darkening, starspots, and the shape of the star and the occulting body should be carried out.  Finally, my estimates of the occultation rates are very crude; a more advanced model of the stellar population as in \citet{Cooray03-Rates} would be useful.

\section*{Acknowledgements}
I thank Scott Tremaine for a reading of this paper and comments{\Blue, and Wystan Benbow for additional discussion.  I thank the referee for their useful comments.}  I {\Blue was} supported by a Jansky Fellowship from the National Radio Astronomy Observatory.  The National Radio Astronomy Observatory is operated by Associated Universities, Inc., under cooperative agreement with the National Science Foundation.

\appendix

\section{Precision needed for numerical computation of light curves}
\label{sec:Precision}
\emph{Poisson's Spot} -- No matter how deep the rest of the transit is, the diffraction pattern $I_{\lambda}$ of a circular screen against a point source always reaches 1 at the exact centre, a phenomenon known as Poisson's spot.  When the screen is much smaller than the Fresnel scale, the intensity slowly oscillates away from the centre.  However, when the screen is much bigger than the Fresnel scale, the diffraction pattern is very nearly a geometrical shadow.  In this case, Poisson's spot is an extremely narrow spike in the centre surrounded by darkness.

When $0 < r_0 \ll \rho$, $J_n \approx (\pi r_0 \rho / 2)^n / n!$.  One can then show that the diffraction pattern very near the centre is 
\begin{equation}
I_{\lambda} \approx [J_0 (\pi r_0 \rho)]^2.
\end{equation}
This function starts out at 1 for $r_0 = 0$ and oscillates around zero with decreasing amplitude, with the first zero at $\pi r_0 \rho \approx 2$.  In order to resolve the behaviour of Poisson's spot, the resolution near zero must be smaller than $\sim 2 / (\pi \rho)$.  The largest $\rho$ of the models I have considered is for a 1 km radius object 10 AU away, where $\rho = 2.1$.  Thus, the resolution of $\Delta r_0 = 0.001$ is sufficient to resolve Poisson's spot in all models considered here.

\emph{The edge of the shadow of large objects} -- The edge of the shadow of objects larger than the Fresnel scale is quite thin.  This poses two obstacles for computing the thickness of the shadow.  First, the series used to calculate the Lommel functions at $r_0 = \rho$ converges very slowly.  The Lommel functions, as defined in \citet{Roques87}, are
\begin{equation}
\label{eqn:LommelDef}
U_n (x, y) = \sum_{k = 0}^{\infty} (-1)^k \left(\frac{x}{y}\right)^{n + 2k} J_{n + 2k} (\pi x y),
\end{equation}
where $J_{n + 2k}$ is a Bessel function of the first kind.  From equation~\ref{eqn:LommelDef}, we have:
\begin{equation}
U_n (x, x) = \sum_{k = 0}^{\infty} (-1)^k J_{n + 2k} (\pi x^2).
\end{equation}
The Bessel functions $J_i (y)$ rise from zero at $y = 0$ to a peak near $y = i$, and {\Blue then} oscillate with a slowly declining amplitude.  Thus, we need to include $j$ terms where $j \gg \pi x^2$.  However, the series converges rapidly for $r_0$ that are not equal to $\rho$, because of the $(r_0 / \rho)$ ratios in equation~\ref{eqn:LommelDef}.  Furthermore, I calculate the necessary Bessel terms using the GNU Science Library function {\tt gsl\_sf\_bessel\_Jn\_array}, which uses a recurrence relation that starts at the $J_i$ with the greatest $i$ and moves downward to $J_0$.  Yet the values of $J_i$ are so near to zero for $r_0 \ll \rho$ that the function {\Blue can} underflow, so we have to make sure that we start from some $J_i(y)$ that is much greater than the minimum value of {\tt double} precision numbers, $\sim 10^{-300}$.  

For a desired precision $\ll 1$, we can set an upper bound on the maximum $i$ of Bessel terms that we need using the following approximation:
\begin{equation}
\label{eqn:JApprox}
J_i (y) \approx \frac{(y / 2)^{i}}{\Gamma(i + 1)}.
\end{equation}
Using Newton's method, I solve for the $i$ where this approximation gives $J_i(y) \approx 10^{-50}$; the remaining terms are then insignificant.

Unfortunately, equation~\ref{eqn:JApprox} is only an upper bound, and the estimated $J_i (y)$ becomes much larger than the actual value when $y \gg \sqrt{i + 1}$.  Thus, there is still a risk of underflow when using this approximation.  Fortunately, the largest $\rho$ value in my models is for a 1 km radius object at 10 AU at $3000\ \Ang$, where $\rho = 2.1$ and $\pi \rho^2 = 14.0$.  According to~\ref{eqn:JApprox}, I would then need to calculate up to $J_{79}$ at $x = 14.0$.  The actual value of $J_{79} (14) = 6.4 \times 10^{-51}$, so there is no risk of underflow.  

Beyond the shadow's edge, $\pi \rho r_0$ is even larger, and even higher order Bessel functions have significant values.  However, I cut off the series expansion at $J_{99} (y)$.  This has no effect on the calculated diffraction patterns: the $j$th term is $(r_0 / \rho)^{-j} J_j(\pi \rho r_0) \la (\pi \rho^2 / 2) / \Gamma(j + 1)$.  Thus, these terms are no larger than when $r_0 = \rho$.  

Second, we need to be sure our spacing in $r_0$ resolves the shadow's edge.  For $\rho \gg 1$, $I_{\lambda}$ transitions from nearly 0 to nearly 1 in a very narrow interval.  Yet $\pi r_0 \rho$ changes slowly, so the edge does not arise because of the Bessel functions, but from the $(r_0 / \rho)$ terms being taken to a high power.  When $r_0 = \rho$, the largest term in the Lommel function is where $i \approx \pi r_0 \rho = \pi \rho^2$.  The width of the shadow's edge is roughly defined by where this term is suppressed:
\begin{align}
1 \approx |(1 \pm \Delta r_0)^{\pi \rho^2}| - 1 \approx \pi \rho^2 |\Delta r_0|.
\end{align}
Thus, the model needs a resolution of $|\Delta r_0| \la 1 / (\pi \rho^2)$.  This is easily fulfilled by my models, since the largest $\pi \rho^2 = 14.0$ and I use steps of $dr_0 = 0.001$.

\emph{The distant fringes of the diffraction pattern} -- The calculation of the diffraction pattern of objects much smaller than the Fresnel scale {\Blue presents} its own complications.  In this case, when $r_0 \gg \rho$, the diffraction pattern has the form:
\begin{equation}
I_{\lambda} \approx 1 - \pi \rho^2 \sin \left[\frac{\pi}{2}(r_0^2 + \rho^2)\right]
\end{equation}
when $r_0 \la 1 / (\pi \rho)$.  The smallest considered $\rho$ in this paper is a 10 meter radius object at $10^4$ AU and $\lambda = 6000\ \Ang$.  Then $\rho = 0.0003$ and the approximation is valid to $r_0 \approx 1000$, far beyond the maximum $r_0$ of 15 that I considered.  The monochromatic diffraction pattern of one of these objects occulting a point source is essentially a ringing of constant amplitude but with the period growing shorter the further from the centre.  The period of the ringing is $P \approx \pi / r_0$.  Note that the partial derivatives of $I_{\lambda}$ with $r_0$ and $\rho$ have the same periods.  

What is actually measured, though, is these ringing patterns integrated over wavelength and a star's disc.  Since the Fresnel scale changes with wavelength, the phases of the diffraction pattern at two wavelengths differ at large $r_0$, resulting in interference and suppressing $I_{\rm point}$.  Likewise, when the width of the star's disc is greater than $P$, the smearing of the diffraction pattern over different phases suppresses $I_{\star}$.  As the wavelengths of photons detected by the PMT detectors of IACTs can vary by a factor $6000\ \Ang / 3000\ \Ang = 2$, it is clear that the interference from the different patterns at these wavelengths suppresses the intensity of the diffraction pattern early on, at $r_0 \approx \pi / 2$, no matter what $\rho$ is.  So most of the power in the observed patterns is near the pattern's centre.

Although the signal observed by PMTs is suppressed, it does not fall to zero immediately.  The sensitivity of IACTs is great enough to actually detect some of the signal even in these regions (see Figure~\ref{fig:CrossSections}).  When integrating over wavelength and over the star's disc, the phase of the ringing must change by less than $\sim 1$ with each step in order to properly compute the amount of interference.  Failing to resolve the phase within the pattern results in spurious ``noise'' in the resulting diffraction pattern.  This noise implies there is a signal where there is none, leading to incorrect detections.  

When calculating $I_{\rm point}$, the Fresnel scale changes from $r_0$ to $r_0 - dr_0$ as wavelength integrates from $\lambda$ to $\lambda + d\lambda$.  Specifically, $dr_0 = r_0 d\lambda / (2\lambda)$.  To resolve the interference correctly, we need $dr_0 \ll P / (2 \pi)$.  Thus, the necessary spectral resolution is
\begin{equation}
d\lambda \ll \lambda / r_0^2.
\end{equation}
In my models, I conservatively use a wavelength step of $d\lambda = 0.1 \lambda_{\rm min} / (\pi {\rm max}[r_0]^2)$.

Likewise, when integrating over the stellar disc (equation~\ref{eqn:IStarIntegral}), $dr_0 \ll P / (2 \pi)$:
\begin{equation}
dr_0^{\prime} \ll \frac{1}{r_0}.
\end{equation}
The necessary resolution in my models is $1/15$; the actual step size of $dr_0^{\prime} = 0.001$ is much smaller than this.  I conclude that my model light curves have the correct amount of interference.

Note that calculating the diffraction pattern at high $r_0$ takes ${\cal O}(r_0^2)$ time, since the step size must decrease inversely with distance.

\section{Derivation of Derivatives}
\label{sec:Derivations}
Taking advantage of the identities for the Bessel functions \citep{Abramowitz65}, 
\begin{align}
\frac{dJ_{\nu} (z)}{dz} & = \frac{1}{2} [J_{\nu - 1} (z) - J_{\nu + 1} (z)] \\
J_{\nu} (z) & = \frac{z}{2i} (J_{\nu - 1} (z) + J_{\nu + 1} z)\\ 
J_1 (z) & = -J_{-1} (z),
\end{align}
it can be shown that the derivatives of the Lommel functions (equation~\ref{eqn:LommelDef}) are
\begin{align}
\frac{\partial U_n (x, y)}{\partial x} & = \pi x U_{n - 1} (x, y)\\
\frac{\partial U_n (x, y)}{\partial y} & = -\pi y U_{n + 1} (x, y)
\end{align}
for $n \ge 0$.  

The derivatives of $I_{\lambda} (r)$ are thus:
\begin{equation}
\frac{\partial I_{\lambda}}{\partial r} = \left\{ \begin{array}{ll} 
						\displaystyle -2 \pi r \Bigl[U_2 (\rho, r) U_3 (\rho, r) + U_1 (\rho, r) U_2 (\rho, r) \Bigr. & \\
                                                \displaystyle \left. + \cos\left(\frac{\pi}{2}(r^2 + \rho^2)\right) \left(U_3 (\rho, r) + U_1 (\rho, r)\right)\right] & (r \ge \rho) \\
                                                \\
						2 \pi r \left[U_0 (r, \rho) U_{-1} (r, \rho) + U_0 (r, \rho) U_1 (r, \rho)\right] & (r \le \rho) \end{array} \right.
\end{equation}
and
\begin{equation}
\frac{\partial I_{\lambda}}{\partial \rho} = \left\{ \begin{array}{ll} 
						\displaystyle 2 \pi \rho \Bigl[U_2 (\rho, r) U_1 (\rho, r) + U_1 (\rho, r) U_0 (\rho, r) \Bigr. & \\
                                                \displaystyle \left. + \sin\left(\frac{\pi}{2}(r^2 + \rho^2)\right) \left(U_2 (\rho, r) + U_0 (\rho, r)\right)\right] & (r \ge \rho) \\
                                                \\
						-2 \pi r \left[U_0 (r, \rho) U_1 (r, \rho) + U_1 (r, \rho) U_2 (r, \rho)\right] & (r \le \rho). \end{array} \right.
\end{equation}

\end{document}